\documentclass[seceq]{ptptex}
\usepackage{wrapft}
\usepackage{graphicx}

\newcommand{\bra}[1]{\langle {#1} |}     
\newcommand{\ket}[1]{| {#1} \rangle}     

\newcommand{\ovl}[1]{\overline{#1}}

\def\beq{\begin{eqnarray}}
\def\eeq{\end{eqnarray}}
\def\bsub{\begin{subequations}}
\def\esub{\end{subequations}}
\def\b{\begin{equation}}
\def\bs{\begin{split}}
\def\es{\end{split}}
\def\e{\end{equation}}

\begin{document}

\title{Spin Polarization versus Color-Flavor-Locking in\\
High Density Quark Matter}

\author{Yasuhiko {\sc Tsue}$^{1,2}$, {Jo\~ao da {\sc Provid\^encia}}$^{2}$, {Constan\c{c}a {\sc Provid\^encia}}$^{2}$, 
{Masatoshi {\sc Yamamura}}$^{3}$ and {Henrik {\sc Bohr}}$^{4}$
}

\inst{$^{1}${Physics Division, Faculty of Science, Kochi University, Kochi 780-8520, Japan}\\
$^{2}${Center for Computational Physics, Departamento de F\'{i}sica, Universidade de Coimbra, 3004-516 Coimbra, 
Portugal}\\
$^{3}${Department of Pure and Applied Physics, 
Faculty of Engineering Science, Kansai University, Suita 564-8680, Japan}\\
$^{4}${Department of Physics, B.307, Danish Technical University, DK-2800 Lyngby, Denmark}
}

\abst{
It is shown that spin polarization with respect to each flavor in three-flavor quark matter occurs instead of the color-flavor locking 
at high baryon density by using the Nambu-Jona-Lasinio model with four-point tensor-type interaction. 
Also, it is indicated that the order of phase transition between the color-flavor locked phase and the spin polarized phase 
is the first order by means of the second order perturbation theory.
}


\maketitle

\section{Introduction}

One of recent interests about the physics governed by the quantum chromodynamics (QCD) may be to understand 
the structure of phase diagram on a plane with respect to, for example, temperature and baryon chemical potential or  external magnetic field, isospin chemical potential 
and so forth \cite{FH}. 
Especially, under extreme conditions such as high baryon density, it is interesting what phase is favorable and is realized. 
In the region with high baryon density and low temperature in the quark matter, it is believed that 
there exists the two-flavor color superconducting (2SC) phase or the color-flavor locked (CFL) phase \cite{CS}. 
In the preceding study, it was indicated that the spin polarized phase may appear at high baryon density due to 
the pseudovector-type interaction between quarks \cite{Tatsumi}. 
However, in the limit of the quark mass being zero, it has been shown that the spin polarized phase disappears \cite{Maedan}. 

In our preceding paper \cite{oursPTP}, it was shown that the quark spin polarized (SP) phase in two-flavor case is realized in the region of high baryon density 
by the use of the Nambu-Jona-Lasinio (NJL) model \cite{NJL} devised by four-point tensor-type interaction with chiral symmetry \cite{arXiv}. 
Further, since 2SC phase may exist in two-flavor QCD at high baryon density, it is also investigated whether the quark spin polarized phase 
is realized or not against the 2SC phase. 
As a result, it was shown that the quark spin polarized phase is actually realized in the same two-flavor NJL model adding to 
the quark-pair interaction \cite{oursPTEP}. 

In this paper, a possibility of quark spin polarization for each flavor is investigated in the three-flavor case by using the NJL model 
with four-point tensor-type interaction. 
In three-flavor case, CFL phase may be realized at high baryon density. 
Thus, the quark-pairing interaction is introduced and it is investigated which phase, namely CFL phase or SP phase, is favorable energetically and 
is realized at high baryon density. 
Also, if both phases exist, it is necessary to discuss the order of phase transition between CFL and SP phase. 

This paper is organized as follows: 
In the next section, the NJL-type model Hamiltonian is explained. 
In \S 3 with Appendix A, under the above-derived Hamiltonian, the CFL phase without the condensate of the quark spin polarization and/or the SP 
phase without color superconducting gap are discussed. 
In \S 4, numerical results are given and the realized phase in certain density regions is considered. 
In \S 5, based on the CFL phase and dealing with tensor-type interaction as a perturbation term, 
the order of phase transition from CFL to SP phase is discussed. 
Some expressions needed in the calculation of second order perturbation are given in Appendix B. 
The last section is devoted to a summary and concluding remarks.

\setcounter{equation}{0}

\section{Hamiltonian showing three-flavor 
color superconductivity and spin polarization based on the NJL model}

Let us start with the following NJL type Lagrangian density: 
\beq\label{1}
& &{\cal L}={\bar \psi}i\gamma^\mu\partial_\mu\psi
-\frac{G}{4}({\bar \psi}\gamma^{\mu}\gamma^{\nu}\lambda_{k}^f\psi)
({\bar\psi}\gamma_{\mu}\gamma_{\nu}\lambda_{k}^f\psi)
+\frac{G_c}{2}({\bar \psi}i\gamma_5\lambda_{a}^c\lambda_{k}^f\psi^C)
({\bar\psi}^Ci\gamma_5\lambda_{a}^c\lambda_{k}^f\psi)\ ,  \qquad
\eeq
where $\psi^C=C{\bar \psi}^T$ with $C=i\gamma^2\gamma^0$ being the charge conjugation operator. 
Also, $\lambda_{k}^f$ and $\lambda_{a}^c$ are the flavor and color $su(3)$ 
Gell-Mann matrices, respectively. 
Here, the NJL Lagrangian density contains other four point interaction parts 
which are not explicitly shown 
such as $G_0 ({\bar \psi}\psi)({\bar \psi}\psi)$ and is invariant under 
chiral transformation. 
However in this paper, some terms are omitted because we investigate only the 
condensates with respect to color-flavor-locking and the each quark-spin polarization in high density 
quark matter in the mean field approximation. 
For example, at high baryon density, the chiral condensate $\langle {\bar \psi}\psi\rangle$ 
is equal to zero. 
Further, in three-flavor case, it is well known that the Kobayashi-Maskawa-'t Hooft (KMT) term \cite{KMtH}
appears, which describes the $U_A(1)$-anomaly and is represented by the 
six-point interaction with determinant-type form in the NJL model. 
However, hereafter, since we adopt the mean field approximation, the KMT term only gives contributions as 
$\langle {\bar \psi}_u\psi_u\rangle\langle {\bar \psi}_d\psi_d\rangle
\langle {\bar \psi}_s\psi_s\rangle$, 
$\langle {\bar \psi}_d\psi_d\rangle
\langle {\bar \psi}_s\psi_s\rangle ({\bar \psi}_u\psi_u)$ and so on, where 
$\psi_f$ represents the quark field with flavor $f$ and 
$\langle \cdots \rangle$ represents the condensate. 
Thus, at high baryon density under investigation in this paper, 
there is no contribution of the KMT term because 
the chiral condensate $\langle{\bar \psi}_f \psi_f\rangle$ is zero. 
As for the quark masses, although the strange quark mass with 0.1 GeV is certainly 
non-zero compared with the up and down quark masses, we may safely ignore the strange 
quark mass at high baryon density with the quark chemical potential being 0.4 - 0.5 GeV under 
consideration in this paper. 
Thus, we ignore the quark mass term in (\ref{1}).

Within the mean field approximation, the above Lagrangian density is expressed as 
\beq\label{2}
& &{\cal L}^{MF}={\bar\psi}i\gamma^\mu\partial_{\mu}\psi
+{\cal L}_T^{MF}+{\cal L}_c^{MF}\ , \nonumber\\
& &\ {\cal L}_T^{MF}=-\sum_{k=3,8}
F_k({\bar \psi}\Sigma_3\lambda_k^f\psi)-\frac{1}{2G}\sum_{k=3,8}F_k^2\ , \nonumber\\
& &\ \ \Sigma_3=-i\gamma^1\gamma^2=
\left(
\begin{array}{cc}
\sigma_3 & 0 \\
0 & \sigma_3 
\end{array}\right)\ , \nonumber\\
& &\ \ F_3=-G\langle {\bar \psi}\Sigma_3\lambda_3^f\psi\rangle\ , \qquad F_8=-G\langle {\bar \psi}\Sigma_3\lambda_8^f\psi\rangle\ , \nonumber\\
& &\qquad {\cal F}_u=F_3+\frac{1}{\sqrt{3}}F_8\ , \quad {\cal F}_d=-F_3+\frac{1}{\sqrt{3}}F_8\ , \quad 
{\cal F}_s=-\frac{2}{\sqrt{3}}F_8\ , \nonumber\\
& &\ {\cal L}_c^{MF}=-\frac{1}{2}\sum_{(a,k)= \{2,5,7\}}
\left(\left(\Delta_{ak}^*({\bar\psi}^Ci\gamma_5\lambda_a^c\lambda_{k}^f\psi)+h.c.
\right)+\frac{1}{2G_C}|\Delta_{ak}|^2\right)\ , \nonumber\\
& &
\ \ 
\Delta_{ak}=-G_c\langle {\bar \psi}^Ci\gamma_5\lambda^c_a\lambda^f_k\psi\rangle\ , 
\eeq
where $h.c.$ represents the Hermitian conjugate term of the preceding one. 
Here, we used the Dirac representation for the Dirac gamma matrices and $\sigma_3$ 
represents the third component of the $2\times 2$ Pauli spin matrices. 
The symbol $\langle\cdots \rangle$ represents the expectation value with respect to a vacuum state. 
The expectation values $F_3$ and $F_8$ correspond 
to the order parameter of the spin alignment which may lead to quark 
spin polarization. 
The expectation value $\Delta_{ak}$ corresponds to the quark-pair condensate which means the existence of the color superconducting phase 
if $\Delta_{ak}\neq 0$.

The mean field Hamiltonian density with quark chemical potential $\mu$ is easily obtained as 
\beq\label{3}
& &{\cal H}_{MF}-\mu{\cal N}={\cal K}_{0}+{\cal H}_{T}^{MF}+{\cal H}_c^{MF}\ , \nonumber\\
& &\quad {\cal K}_{0}={\bar \psi}(-i{\mib \gamma}\cdot{\mib \nabla}-\mu\gamma_0)\psi\ , \nonumber\\
& &\quad {\cal H}_T^{MF}=-{\cal L}_T^{MF}\ , \qquad
{\cal H}_c^{MF}=-{\cal L}_c^{MF}
\eeq
with ${\cal N}=\psi^{\dagger}\psi$. 
In the Dirac representation for the Dirac gamma matrices, the Hamiltonian matrix 
of the spin polarization part in $H_{MF}^{SP}=\int d^3{\mib x}\ ({\cal K}_0+{\cal H}_T^{MF})$ 
is written as 
\beq\label{4}
h_{MF}^{SP}&=&{\mib p}\cdot{\mib \alpha}+{\cal F}_\tau\beta \Sigma_3\nonumber\\
&=&
\left(
\begin{array}{cccc}
{\cal F}_\tau & 0 & p_3 & p_1-ip_2 \\
0 & -{\cal F}_\tau & p_1+ip_2 & p_3 \\
p_3 & p_1-ip_2 & -{\cal F}_\tau & 0 \\
p_1+ip_2 & -p_3 & 0 & {\cal F}_\tau 
\end{array}
\right)\ , 
\eeq 
where $\alpha^i=\gamma^0\gamma^i$ and $\beta=\gamma^0$. 
Here, we define ${\cal F}_{\tau}$ as
\beq\label{5}
{\cal F}_{\tau}=\sum_{k=3,8}F_{k}\lambda_k^f=\left(
F_3+\frac{1}{\sqrt{3}}F_8\right)\delta_{\tau u}
+\left(-F_3+\frac{1}{\sqrt{3}}F_8\right)\delta_{\tau d}
-\frac{2}{\sqrt{3}}F_8\delta_{\tau s}\ .
\eeq
For good helicity states, this Hamiltonian matrix is easily diagonalized in the case ${\cal F}_{\tau}=0$. 
For simplicity, we rotate around $p_3$ axis and we set $p_2=0$ without loss of generality. 
In this case, we derive $\kappa=U^{-1}h_{MF}^{SP}U$ as follows: 
\beq\label{6}
U&=&
\frac{1}{2\sqrt{p}}
\left(
\begin{array}{cccc}
\sqrt{p+p_3} & \sqrt{p-p_3} & -\sqrt{p+p_3} & -\sqrt{p-p_3} \\
\frac{p_1}{|p_1|}\sqrt{p-p_3} & -\frac{p_1}{|p_1|}\sqrt{p+p_3} & -\frac{p_1}{|p_1|}\sqrt{p-p_3} & \frac{p_1}{|p_1|}\sqrt{p+p_3} \\
\sqrt{p+p_3} & -\sqrt{p-p_3} & \sqrt{p+p_3} & -\sqrt{p-p_3} \\
\frac{p_1}{|p_1|}\sqrt{p-p_3} & \frac{p_1}{|p_1|}\sqrt{p+p_3} & \frac{p_1}{|p_1|}\sqrt{p-p_3} & \frac{p_1}{|p_1|}\sqrt{p+p_3} \\
\end{array}
\right)\ , \nonumber\\
\kappa&=&U^{-1}h_{MF}^{SP}U \nonumber\\
&=&\left(
\begin{array}{cccc}
p & 0 & 0 & 0 \\
0 & p & 0 & 0 \\
0 & 0 & -p & 0 \\
0 & 0 & 0 & -p \\
\end{array}
\right)+
\frac{{\cal F}_\tau}{p}
\left(
\begin{array}{cccc}
0  & |p_1| & -p_3  & 0 \\
|p_1| & 0 & 0 & p_3 \\
-p_3 & 0 & 0 & |p_1| \\
0 & p_3 & |p_1| & 0 
\end{array}
\right)\ . 
\eeq
Finally, in the original basis rotated around $p_3$-axis, $|p_1|$ is replaced to $\sqrt{p_1^2+p_2^2}$. 
Thus, the many-body Hamiltonian can be expressed by means of the quark creation and 
annihilation operators as is seen later.

As for the Hamiltonian matrix of the color superconducting part, $H_c^{MF}=\int d^3{\mib x}{\cal H}_c^{MF}$, 
we can derive another expression by using the quark-creation and annihilation operators with respect to good helicity states, similarly to Ref.\cite{oursPTEP}. 
As a result, 
in the basis of good helicity states, the relevant combination of the mean field Hamiltonian 
$H_{MF}=\int d^3{\mib x}{\cal H}_{MF}$ and the quark number $N=\int d^3{\mib x}{\cal N}$ is given by 
\beq\label{7}
& &H=H_0-\mu {\hat N}+V_{\rm SP}+V_{\rm CFL}+V\cdot\frac{1}{2G}(F_3^2+F_8^2)+V\cdot\frac{3\Delta^2}{2G_c}\ , \nonumber\\
& &H_{0}-\mu {\hat N}=
\sum_{{\bf\small p}\eta\tau\alpha}\left[
(|{\mib p}|-\mu)c_{{\bf\small p}\eta\tau\alpha}^{\dagger}c_{{\bf\small p}\eta\tau\alpha}
-(|{\mib p}|+\mu){\tilde c}_{{\bf\small p}\eta\tau\alpha}^{\dagger}{\tilde c}_{{\bf\small p}\eta\tau\alpha}^{\dagger}\right]\ , \nonumber\\
& &V_{\rm SP}=
\sum_{{\bf\small p}\eta\tau\alpha}{\cal F}_{\tau}\biggl[
\frac{\sqrt{p_1^2+p_2^2}}{|{\mib p}|}(c_{{\bf\small p}\eta\tau\alpha}^{\dagger}c_{{\bf\small p}-\eta\tau\alpha}
+{\tilde c}_{{\bf\small p}\eta\tau\alpha}^{\dagger}{\tilde c}_{{\bf\small p}-\eta\tau\alpha})
\nonumber\\
& &\qquad\qquad\qquad\qquad\qquad\qquad
-\eta\frac{p_3}{|{\mib p}|}(c_{{\bf\small p}\eta\tau\alpha}^{\dagger}{\tilde c}_{{\bf\small p}\eta\tau\alpha}
+{\tilde c}_{{\bf\small p}\eta\tau\alpha}^{\dagger}c_{{\bf\small p}\eta\tau\alpha})\biggl]\ , \nonumber\\
& &V_{\rm CFL}=\frac{\Delta}{2}\sum_{{\bf\small p}\eta}\sum_{\alpha\alpha'\alpha''}
\sum_{\tau\tau'}
(c_{{\bf\small p}\eta\alpha\tau}^{\dagger}c_{-{\bf\small p}\eta\alpha'\tau'}^{\dagger}
+c_{-{\bf\small p}\eta\alpha'\tau'}c_{{\bf\small p}\eta\alpha\tau}
\nonumber\\
& &\qquad\qquad\qquad\qquad\qquad\ 
+{\tilde c}_{{\bf\small p}\eta\alpha\tau}^{\dagger}{\tilde c}_{-{\bf\small p}\eta\alpha'\tau'}^{\dagger}
+{\tilde c}_{-{\bf\small p}\eta\alpha'\tau'}{\tilde c}_{{\bf\small p}\eta\alpha\tau})
\epsilon_{\alpha\alpha'\alpha''}\epsilon_{\tau\tau'\tau_{\alpha''}}\phi_p
\ , 
\eeq
where $V$ represents the volume in the box normalization.
Here, $c^{\dagger}_{{\mib p}\eta\tau\alpha}$ and ${\tilde c}^{\dagger}_{{\mib p}\eta\tau\alpha}$ represent 
the quark and antiquark creation operators
with momentum ${\mib p}$, helicity $\eta=\pm$, flavor index $\tau$ and color $\alpha$.  
It should be noted that $\alpha\ (=1,2,3)$ and $\tau\ (=u,d,s)$ represent color and flavor, respectively, in which especially 
we understand $\tau_1=u$, $\tau_2=d$ and $\tau_3=s$. 
Hereafter, we will use $p\equiv({\mib p},\eta)$ and ${\bar p}\equiv (-{\mib p},\eta)$ 
as abbreviated notations.
Also, $\epsilon_{\tau\tau'\tau''}$ and $\epsilon_{\alpha\alpha'\alpha''}$ represent the complete antisymmetric tensor for 
the flavor and color indices. 
We define $p=\sqrt{p_1^2+p_2^2+p_3^2}$, that is, the magnitude of momentum. 
\footnote{
Here, it is necessary to introduce $\phi_p$ in (\ref{7}) 
for the character of fermion operator $c_{{\bf\small p}\eta\tau\alpha}$ and ${\tilde c}_{{\bf\small p}\eta\tau\alpha}$. 
Namely, for example, 
\beq
\sum_{{\small\bf p}\eta\{\alpha\}\{\tau\}}\!\!\!\!
c^{\dagger}_{{\small\bf p}\eta\tau\alpha}c^{\dagger}_{-{\small\bf p}\eta\tau'\alpha'}\phi_p\epsilon_{\alpha\alpha'\alpha''}
\epsilon_{\tau\tau'\tau_{\alpha''}}
&=&\frac{1}{2}\!\!\!\!
\sum_{{\small\bf p}\eta\{\alpha\}\{\tau\}}(c^{\dagger}_{{\small\bf p}\eta\tau\alpha}c^{\dagger}_{-{\small\bf p}\eta\tau'\alpha'}\phi_p
+c^{\dagger}_{-{\small\bf p}\eta\tau\alpha}c^{\dagger}_{{\small\bf p}\eta\tau'\alpha'}\phi_{\bar p})\epsilon_{\alpha\alpha'\alpha''}
\epsilon_{\tau\tau'\tau_{\alpha''}}
\nonumber\\
&=&\frac{1}{2}\!\!\!\!
\sum_{{\small\bf p}\eta\{\alpha\}\{\tau\}}(c^{\dagger}_{{\small\bf p}\eta\tau\alpha}c^{\dagger}_{-{\small\bf p}\eta\tau'\alpha'}\phi_p
-c^{\dagger}_{{\small\bf p}\eta\tau'\alpha'}c^{\dagger}_{-{\small\bf p}\eta\tau\alpha}\phi_{\bar p})\epsilon_{\alpha\alpha'\alpha''}
\epsilon_{\tau\tau'\tau_{\alpha''}}
\nonumber\\
&=&\frac{1}{2}\!\!\!\!
\sum_{{\small\bf p}\eta\{\alpha\}\{\tau\}}(c^{\dagger}_{{\small\bf p}\eta\tau\alpha}c^{\dagger}_{-{\small\bf p}\eta\tau'\alpha'}\phi_p
-c^{\dagger}_{{\small\bf p}\eta\tau\alpha}c^{\dagger}_{-{\small\bf p}\eta\tau'\alpha'}\phi_{\bar p})\epsilon_{\alpha\alpha'\alpha''}
\epsilon_{\tau\tau'\tau_{\alpha''}}\ . \nonumber
\eeq
Thus, we find $\phi_p=-\phi_{\bar p}$ with $p=({\mib p},\eta)$ and ${\bar p}=(-{\mib p},\eta)$. 
}
In (\ref{7}), the color and flavor are locked due to $V_{\rm CFL}$. 
Namely, the combination $\langle c_{{\bar p} 1 u}c_{p 2 d}\rangle
-\langle c_{{\bar p} 1 d}c_{p 2 u}\rangle$, 
$\langle c_{{\bar p} 2 d}c_{p 3 s}\rangle
-\langle c_{{\bar p} 2 s}c_{p 3 d}\rangle$ and 
$\langle c_{{\bar p} 3 s}c_{p 1 u}\rangle
-\langle c_{{\bar p} 3 u}c_{p 1 s}\rangle$
appear in the CFL condensate $\Delta$, 
in which we define 
\beq\label{8}
& &
\Delta_{\alpha''\tau_{\alpha''}}=G_c\sum_{{\small\bf p}\eta\alpha\alpha'}\sum_{\tau\tau'}
\langle c_{-{\small\bf p}\eta\alpha'\tau'} c_{{\small\bf p}\eta\alpha\tau}\rangle
\epsilon_{\alpha\alpha'\alpha''}\epsilon_{\tau\tau'\tau_{\alpha''}}\phi_p\ , \nonumber\\
& &\ \ \Delta=\Delta_{1u}=\Delta_{2d}=\Delta_{3s}\ .
\eeq
The symmetry $su(3)_{\rm CFL}$ remains because the symmetry breaking pattern is 
$su(3)_c\otimes su(3)_f \rightarrow su(3)_{\rm CFL}$.

\setcounter{equation}{0}

\section{Color-flavor locked phase without spin polarization 
and spin polarized phase without color-flavor-locking}

\subsection{Color-flavor locked phase without spin polarization}

Let us consider the case $F_k=0$, which leads to the color superconductor without spin polarization \cite{2,3}. 
The Hamiltonian is expressed as 
\beq\label{9}
& &H_{\rm eff}=H_{\rm 0}-\mu {\hat N}+V_{\rm CFL}+V\cdot\frac{3\Delta^2}{2G_c}\ , \nonumber\\
& &H_{\rm 0}-\mu {\hat N}=
\sum_{{\bf\small p}\eta\tau\alpha}\left[
(|{\mib p}|-\mu)c_{{\bf\small p}\eta\tau\alpha}^{\dagger}c_{{\bf\small p}\eta\tau\alpha}
-(|{\mib p}|+\mu){\tilde c}_{{\bf\small p}\eta\tau\alpha}^{\dagger}{\tilde c}_{{\bf\small p}\eta\tau\alpha}^{\dagger}\right]\ , \nonumber\\
& &V_{\rm CFL}=\frac{\Delta}{2}\sum_{{\bf\small p}\eta}\sum_{\alpha\alpha'\alpha''}
\sum_{\tau\tau'}
(c_{{\bf\small p}\eta\alpha\tau}^{\dagger}c_{-{\bf\small p}\eta\alpha'\tau'}^{\dagger}
+c_{-{\bf\small p}\eta\alpha'\tau'}c_{{\bf\small p}\eta\alpha\tau}
\nonumber\\
& &\qquad\qquad\qquad\qquad\qquad\quad  
+{\tilde c}_{{\bf\small p}\eta\alpha\tau}^{\dagger}{\tilde c}_{-{\bf\small p}\eta\alpha'\tau'}^{\dagger}
+{\tilde c}_{-{\bf\small p}\eta\alpha'\tau'}{\tilde c}_{{\bf\small p}\eta\alpha\tau})
\epsilon_{\alpha\alpha'\alpha''}\epsilon_{\tau\tau'\tau_{\alpha''}}\phi_p
\ . 
\eeq
Hereafter, we use an abbreviated notation $p=({\mib p}, \eta)$ and ${\bar p}=(-{\mib p},\eta)$. 
The commutation relations are calculated as 
\beq\label{10}
& &[\ H_{\rm eff}\ , \ c_{p1u}\ ]=-(\epsilon_{pu}-\mu)c_{p1u}-\Delta(c_{{\bar p}2d}^{\dagger}+c_{{\bar p}3s}^{\dagger})\phi_p\ , \nonumber\\
& &[\ H_{\rm eff}\ , \ c_{p2u}\ ]=-(\epsilon_{pu}-\mu)c_{p2u}+\Delta c_{{\bar p}1d}^{\dagger}\phi_p\ , \nonumber\\
& &[\ H_{\rm eff}\ , \ c_{p3u}\ ]=-(\epsilon_{pu}-\mu)c_{p3u}+\Delta c_{{\bar p}1s}^{\dagger}\phi_p\ , \nonumber\\
& &[\ H_{\rm eff}\ , \ c_{p1d}\ ]=-(\epsilon_{pd}-\mu)c_{p1d}+\Delta c_{{\bar p}2u}^{\dagger}\phi_p\ , \nonumber\\
& &[\ H_{\rm eff}\ , \ c_{p2d}\ ]=-(\epsilon_{pd}-\mu)c_{p2d}-\Delta(c_{{\bar p}3s}^{\dagger}+c_{{\bar p}1u}^{\dagger})\phi_p\ , \nonumber\\
& &[\ H_{\rm eff}\ , \ c_{p3d}\ ]=-(\epsilon_{pd}-\mu)c_{p3d}+\Delta c_{{\bar p}2s}^{\dagger}\phi_p\ , \nonumber\\
& &[\ H_{\rm eff}\ , \ c_{p1s}\ ]=-(\epsilon_{ps}-\mu)c_{p1s}+\Delta c_{{\bar p}3u}^{\dagger}\phi_p\ , \nonumber\\
& &[\ H_{\rm eff}\ , \ c_{p2s}\ ]=-(\epsilon_{ps}-\mu)c_{p2s}+\Delta c_{{\bar p}3d}^{\dagger}\phi_p\ , \nonumber\\
& &[\ H_{\rm eff}\ , \ c_{p3s}\ ]=-(\epsilon_{ps}-\mu)c_{p3s}-\Delta(c_{{\bar p}1u}^{\dagger}+c_{{\bar p}2d}^{\dagger})\phi_p\ , 
\eeq
where $\epsilon_{pu}=\epsilon_{pd}=\epsilon_{ps}=|{\mib p}|(=\epsilon_p)$. 
Thus, $(1u)$, $(2d)$ and $(3s)$ become combined with each other. 
Further, the sets $(2u,1d)$, $(3u,1s)$ and $(3d,2s)$ are mixed each other. 
First, let us consider the case $\varepsilon_{p\tau}>\mu$. 
Thus, for example, we are led to consider new operators such as the following one:
\beq\label{11}
d_{p2u}^{\dagger}=X_p c_{p2u}^{\dagger}+Y_p c_{{\bar p}1d}\ . 
\eeq
Here, we demand that this operator should satisfy the following commutation relation in order to diagonalize the Hamiltonian $H_{\rm eff}$: 
\beq\label{12}
[\ H_{\rm eff}\ , \ d_{p2u}^{\dagger}\ ]&=&[(\epsilon_{pu}-\mu)X_p-\Delta Y_p\phi_p]c_{p2u}^{\dagger}+[-(\epsilon_{pd}-\mu)Y_p-\Delta X_p \phi_p]c_{{\bar p}1d}\nonumber\\
&\equiv & \omega\ d_{p2u}^{\dagger}\ . 
\eeq
Then, $X_p$ and $Y_p$ are determined from the following equation:
\beq\label{13}
\left(
\begin{array}{cc}
\epsilon_{p}-\mu & -\Delta \phi_p \\
-\Delta \phi_p & -(\epsilon_{p}-\mu) 
\end{array}
\right)
\left(
\begin{array}{c}
X_p \\ Y_p 
\end{array}\right)
=\omega\left(
\begin{array}{c}
X_p \\ Y_p 
\end{array}\right)\ . 
\eeq 
As a result, we can derive the following:
\beq\label{14}
& &X_p=\frac{1}{\sqrt{2}}\left[1+\frac{{\bar \epsilon}_{p}}{\sqrt{{\bar \epsilon}_{p}^2+\Delta^2}}\right]^{1/2}\ ,\qquad
Y_p=-\frac{1}{\sqrt{2}}\left[1-\frac{{\bar \epsilon}_{p}}{\sqrt{{\bar \epsilon}_{p}^2+\Delta^2}}\right]^{1/2}
\phi_p\ , \\
& &\ \ 
\omega_{p}=\sqrt{{\bar \epsilon}_{p}^2+\Delta^2}\ , \qquad
{\bar \epsilon}_{p}=\epsilon_{p}-\mu\ . \nonumber
\eeq
Similarly, we can introduce new operators and the results are summarized as follows for $|{\mib p}|>\mu$: 
\beq\label{15}
& &d_{p;1}^{\dagger}=x^{(1)}_p(c_{p1u}^{\dagger}+c_{p2d}^{\dagger}+c_{p3s}^{\dagger})+y_p^{(1)}(c_{{\bar p}1u}+c_{{\bar p}2d}+c_{{\bar p}3s})\ , \nonumber\\
& &\qquad x^{(1)}_p=\frac{1}{\sqrt{6}}\left[1+\frac{{\bar \epsilon}_p}{\sqrt{{\bar \epsilon}_p^2+4\Delta^2}}\right]^{1/2},,\qquad
y^{(1)}_p=\frac{1}{\sqrt{6}}\left[1-\frac{{\bar \epsilon}_p}{\sqrt{{\bar \epsilon}_p^2+4\Delta^2}}\right]^{1/2}\!\!\!\!\cdot\phi_p\ , \ \ 
\nonumber\\
& &\qquad\qquad
\omega_1=\sqrt{{\bar \epsilon}_p^2+4\Delta^2}\ , \qquad\nonumber\\
& &d_{p;2}^{\dagger}=x^{(2)}_p(c_{p1u}^{\dagger}-c_{p2d}^{\dagger})+y_p^{(2)}(c_{{\bar p}1u}-c_{{\bar p}2d})\ , \nonumber\\
& &\qquad x^{(2)}_p=\frac{1}{2}\left[1+\frac{{\bar \epsilon}_p}{\sqrt{{\bar \epsilon}_p^2+\Delta^2}}\right]^{1/2},\quad
y^{(2)}_p=-\frac{1}{2}\left[1-\frac{{\bar \epsilon}_p}{\sqrt{{\bar \epsilon}_p^2+\Delta^2}}\right]^{1/2}\!\!\!\!\cdot\phi_p\ , \nonumber\\
& &\qquad\qquad
\omega_2=\sqrt{{\bar \epsilon}_p^2+\Delta^2}\ , \nonumber\\
& &d_{p;3}^{\dagger}=x^{(3)}_p(c_{p1u}^{\dagger}+c_{p2d}^{\dagger}-2c_{p3s}^{\dagger})+y_p^{(3)}(c_{{\bar p}1u}+c_{{\bar p}2d}-2c_{{\bar p}3s})\ , \nonumber\\
& &\qquad x^{(3)}_p=\frac{1}{2\sqrt{3}}\left[1+\frac{{\bar \epsilon}_p}{\sqrt{{\bar \epsilon}_p^2+\Delta^2}}\right]^{1/2},\quad
y^{(3)}_p=-\frac{1}{2\sqrt{3}}\left[1-\frac{{\bar \epsilon}_p}{\sqrt{{\bar \epsilon}_p^2+\Delta^2}}\right]^{1/2}\!\!\!\!\cdot\phi_p\ , \nonumber\\
& &\qquad\qquad
\omega_3=\sqrt{{\bar \epsilon}_p^2+\Delta^2}\ , \nonumber\\
& &d_{p;4}=d_{p2u}^{\dagger}=X_p c_{p2u}^{\dagger}+Y_p c_{{\bar p}1d}\ , 
\qquad
d_{p;5}=d_{p1d}^{\dagger}=X_p c_{p1d}^{\dagger}+Y_p c_{{\bar p}2u}\ ,
\nonumber\\ 
& &
d_{p;6}=d_{p3u}^{\dagger}=X_p c_{p3u}^{\dagger}+Y_p c_{{\bar p}1s}\ , 
\qquad
d_{p;7}=d_{p1s}^{\dagger}=X_p c_{p1s}^{\dagger}+Y_p c_{{\bar p}3u}\ , 
\nonumber\\  
& &d_{p;8}=d_{p3d}^{\dagger}=X_p c_{p3d}^{\dagger}+Y_p c_{{\bar p}2s}\ , 
\qquad
d_{p;9}=d_{p2s}^{\dagger}=X_p c_{p2s}^{\dagger}+Y_p c_{{\bar p}3d}\ , 
\nonumber\\  
& &\qquad
X_p=\frac{1}{\sqrt{2}}\left[1+\frac{{\bar \epsilon}_p}{\sqrt{{\bar \epsilon}_p^2+\Delta^2}}\right]^{1/2},\quad
Y_p=-\frac{1}{\sqrt{2}}\left[1-\frac{{\bar \epsilon}_p}{\sqrt{{\bar \epsilon}_p^2+\Delta^2}}\right]^{1/2}\!\!\!\!\cdot\phi_p\ , \nonumber\\
& &\qquad\qquad
\omega=\sqrt{{\bar \epsilon}_p^2+\Delta^2}\ ,
\eeq
where ${\bar \epsilon}_p=|{\mib p}|-\mu$, $y_{{\bar p}}^{(i)}=-y_{p}^{(i)}$ and $Y_{{\bar p}}=-Y_p$. 
Inversely, we can derive the following:
\beq\label{16}
& &c_{p1u}^{\dagger}=x_p^{(1)}d_{p;1}^{\dagger}+x_p^{(2)}d_{p;2}^{\dagger}+x_p^{(3)}d_{p;3}^{\dagger}
-(y_p^{(1)}d_{{\bar p};1}+y_p^{(2)}d_{{\bar p};2}+y_p^{(3)}d_{{\bar p};3})\ , \nonumber\\
& &c_{p2d}^{\dagger}=x_p^{(1)}d_{p;1}^{\dagger}-x_p^{(2)}d_{p;2}^{\dagger}+x_p^{(3)}d_{p;3}^{\dagger}
-(y_p^{(1)}d_{{\bar p};1}-y_p^{(2)}d_{{\bar p};2}+y_p^{(3)}d_{{\bar p};3})\ , \nonumber\\
& &c_{p3s}^{\dagger}=x_p^{(1)}d_{p;1}^{\dagger}-2x_p^{(3)}d_{p;3}^{\dagger}
-(y_p^{(1)}d_{{\bar p};1}-2y_p^{(3)}d_{{\bar p};3})\ , \nonumber\\
& &c_{p2u}^{\dagger}=X_pd_{p2u}^{\dagger}-Y_pd_{{\bar p}1d}\ , \quad
c_{p1d}^{\dagger}=X_pd_{p1d}^{\dagger}-Y_pd_{{\bar p}2u}\ , \quad
c_{p3u}^{\dagger}=X_pd_{p3u}^{\dagger}-Y_pd_{{\bar p}1s}\ , \nonumber\\
& &c_{p1s}^{\dagger}=X_pd_{p1s}^{\dagger}-Y_pd_{{\bar p}3u}\ , \quad
c_{p3d}^{\dagger}=X_pd_{p3d}^{\dagger}-Y_pd_{{\bar p}2s}\ , \quad
c_{p2s}^{\dagger}=X_pd_{p2s}^{\dagger}-Y_pd_{{\bar p}3d}\ . \quad \ \ 
\eeq

As for $\epsilon_p<\mu$ with $\epsilon_p=|{\mib p}|$, we can introduce the new operators 
as is similar to the case of $\epsilon_p>\mu$. 
The new operators, ${\bar d}_{{\bar p};i}$, satisfy the diagonalized commutation relations such as 
\beq\label{17}
[\ H_{\rm eff}\ , \ {\bar d}_{{\bar p};i}\ ]=-\omega_i{\bar d}_{{\bar p};i}\ , 
\eeq 
where $\omega_4=\omega_5=\cdots =\omega_9=\omega$. 
Then, the new operators can be derived as 
\beq\label{18}
& &{\bar d}_{{\bar p};1}={\bar x}^{(1)}_p(c_{p1u}^{\dagger}+c_{p2d}^{\dagger}+c_{p3s}^{\dagger})+{\bar y}_p^{(1)}(c_{{\bar p}1u}+c_{{\bar p}2d}+c_{{\bar p}3s})\ , \nonumber\\
& &\qquad {\bar x}^{(1)}_p=\frac{1}{\sqrt{6}}\left[1-\frac{{\bar \epsilon}_p}{\sqrt{{\bar \epsilon}_p^2+4\Delta^2}}\right]^{1/2},\quad
{\bar y}^{(1)}_p=-\frac{1}{\sqrt{6}}\left[1+\frac{{\bar \epsilon}_p}{\sqrt{{\bar \epsilon}_p^2+4\Delta^2}}\right]^{1/2}\!\!\!\!\cdot\phi_p\ , \nonumber\\
& &\qquad\qquad
\omega_1=\sqrt{{\bar \epsilon}_p^2+4\Delta^2}\ , \nonumber\\
& &{\bar d}_{{\bar p};2}={\bar x}^{(2)}_p(c_{p1u}^{\dagger}-c_{p2d}^{\dagger})+{\bar y}_p^{(2)}(c_{{\bar p}1u}-c_{{\bar p}2d})\ , \nonumber\\
& &\qquad {\bar x}^{(2)}_p=\frac{1}{2}\left[1-\frac{{\bar \epsilon}_p}{\sqrt{{\bar \epsilon}_p^2+\Delta^2}}\right]^{1/2},\quad
{\bar y}^{(2)}_p=\frac{1}{2}\left[1+\frac{{\bar \epsilon}_p}{\sqrt{{\bar \epsilon}_p^2+\Delta^2}}\right]^{1/2}\!\!\!\!\cdot\phi_p\ , \nonumber\\
& &\qquad\qquad
\omega_2=\sqrt{{\bar \epsilon}_p^2+\Delta^2}\ , \nonumber\\
& &{\bar d}_{{\bar p};3}={\bar x}^{(3)}_p(c_{p1u}^{\dagger}+c_{p2d}^{\dagger}-2c_{p3s}^{\dagger})+{\bar y}_p^{(3)}(c_{{\bar p}1u}+c_{{\bar p}2d}-2c_{{\bar p}3s})\ , \nonumber\\
& &\qquad {\bar x}^{(3)}_p=\frac{1}{2\sqrt{3}}\left[1-\frac{{\bar \epsilon}_p}{\sqrt{{\bar \epsilon}_p^2+\Delta^2}}\right]^{1/2},\quad
{\bar y}^{(3)}_p=\frac{1}{2\sqrt{3}}\left[1+\frac{{\bar \epsilon}_p}{\sqrt{{\bar \epsilon}_p^2+\Delta^2}}\right]^{1/2}\!\!\!\!\cdot\phi_p\ , \nonumber\\
& &\qquad\qquad
\omega_3=\sqrt{{\bar \epsilon}_p^2+\Delta^2}\ , \nonumber\\
& &{\bar d}_{{\bar p};4}={\bar d}_{{\bar p}2u}={\bar X}_p c_{p2u}^{\dagger}+{\bar Y}_p c_{{\bar p}1d}\ , 
\qquad
{\bar d}_{{\bar p};5}={\bar d}_{{\bar p}1d}={\bar X}_p c_{p1d}^{\dagger}+{\bar Y}_p c_{{\bar p}2u}\ , 
\nonumber\\ 
& &{\bar d}_{{\bar p};6}={\bar d}_{{\bar p}3u}={\bar X}_p c_{p3u}^{\dagger}+{\bar Y}_p c_{{\bar p}1s}\ , 
\qquad
{\bar d}_{{\bar p};7}={\bar d}_{{\bar p}1s}={\bar X}_p c_{p1s}^{\dagger}+{\bar Y}_p c_{{\bar p}3u}\ , 
\nonumber\\  
& &{\bar d}_{{\bar p};8}={\bar d}_{{\bar p}3d}={\bar X}_p c_{p3d}^{\dagger}+{\bar Y}_p c_{{\bar p}2s}\ , 
\qquad
{\bar d}_{{\bar p};9}={\bar d}_{{\bar p}2s}={\bar X}_p c_{p2s}^{\dagger}+{\bar Y}_p c_{{\bar p}3d}\ , 
\nonumber\\  
& &\qquad
{\bar X}_p=\frac{1}{\sqrt{2}}\left[1-\frac{{\bar \epsilon}_p}{\sqrt{{\bar \epsilon}_p^2+\Delta^2}}\right]^{1/2},\quad
{\bar Y}_p=\frac{1}{\sqrt{2}}\left[1+\frac{{\bar \epsilon}_p}{\sqrt{{\bar \epsilon}_p^2+\Delta^2}}\right]^{1/2}\!\!\!\!\cdot\phi_p\ , \nonumber\\
& &\quad\qquad
\omega=\sqrt{{\bar \epsilon}_p^2+\Delta^2}\ .
\eeq
The following inverse relations are obtained:
\beq\label{19}
& &c_{p1u}^{\dagger}={\bar x}_p^{(1)}{\bar d}_{{\bar p};1}+{\bar x}_p^{(2)}{\bar d}_{{\bar p};2}+{\bar x}_p^{(3)}{\bar d}_{{\bar p};3}
-({\bar y}_p^{(1)}{\bar d}_{{p};1}^{\dagger}+{\bar y}_p^{(2)}{\bar d}_{{p};2}^{\dagger}+{\bar y}_p^{(3)}{\bar d}_{{p};3}^{\dagger})\ , \nonumber\\
& &c_{p2d}^{\dagger}={\bar x}_p^{(1)}{\bar d}_{{\bar p};1}-{\bar x}_p^{(2)}{\bar d}_{{\bar p};2}+{\bar x}_p^{(3)}{\bar d}_{{\bar p};3}
-({\bar y}_p^{(1)}{\bar d}_{{p};1}^{\dagger}-{\bar y}_p^{(2)}{\bar d}_{{p};2}^{\dagger}+{\bar y}_p^{(3)}{\bar d}_{{p};3}^{\dagger})\ , \nonumber\\
& &c_{p3s}^{\dagger}={\bar x}_p^{(1)}{\bar d}_{{\bar p};1}-2{\bar x}_p^{(3)}{\bar d}_{{\bar p};3}
-({\bar y}_p^{(1)}{\bar d}_{{p};1}^{\dagger}-2{\bar y}_p^{(3)}{\bar d}_{{p};3}^{\dagger})\ , \nonumber\\
& &c_{p2u}^{\dagger}={\bar X}_p{\bar d}_{{\bar p}2u}-{\bar Y}_p{\bar d}_{{p}1d}^{\dagger}\ , 
\quad
c_{p1d}^{\dagger}={\bar X}_p{\bar d}_{{\bar p}1d}-{\bar Y}_p{\bar d}_{{p}2u}^{\dagger}\ , 
\quad
c_{p3u}^{\dagger}={\bar X}_p{\bar d}_{{\bar p}3u}-{\bar Y}_p{\bar d}_{{p}1s}^{\dagger}\ , \nonumber\\
& &c_{p1s}^{\dagger}={\bar X}_p{\bar d}_{{\bar p}1s}-{\bar Y}_p{\bar d}_{{p}3u}^{\dagger}\ , \quad
c_{p3d}^{\dagger}={\bar X}_p{\bar d}_{{\bar p}3d}-{\bar Y}_p{\bar d}_{{p}2s}^{\dagger}\ , 
\quad
c_{p2s}^{\dagger}={\bar X}_p{\bar d}_{{\bar p}2s}-{\bar Y}_p{\bar d}_{{p}3d}^{\dagger}\ . \qquad\ \ 
\eeq
By using the above operators, we rewrite $H_{\rm eff}$ as is shown in (\ref{a1}) in appendix. 
Noting $X_{p}$ etc. in (\ref{15}) and (\ref{18}), we finally obtain the following diagonalized many-body Hamiltonian without spin polarization:
\beq\label{20}
H_{\rm eff}&=&
\frac{1}{2}\sum_{p\ (\epsilon_p>\mu)}\left[9{\bar \epsilon}_p-\sqrt{{\bar \epsilon}_p^2+4\Delta^2}-8\sqrt{{\bar \epsilon}_p^2+\Delta^2}\right]
\nonumber\\
& &
+\sum_{p\ (\epsilon_p>\mu)}\left[\sqrt{{\bar \epsilon}_p^2+4\Delta^2}d_{p;1}^{\dagger}d_{p;1}
+\sum_{a=2}^9\sqrt{{\bar \epsilon}_p^2+\Delta^2}d_{p;a}^{\dagger}d_{p;a}\right]\nonumber\\
& &+
\frac{1}{2}\sum_{p\ (\epsilon_p<\mu)}\left[9{\bar \epsilon}_p-\sqrt{{\bar \epsilon}_p^2+4\Delta^2}-8\sqrt{{\bar \epsilon}_p^2+\Delta^2}\right]
\nonumber\\
& &
+\sum_{p\ (\epsilon_p<\mu)}\left[\sqrt{{\bar \epsilon}_p^2+4\Delta^2}{\bar d}_{p;1}^{\dagger}{\bar d}_{p;1}
+\sum_{a=2}^9\sqrt{{\bar \epsilon}_p^2+\Delta^2}{\bar d}_{p;a}^{\dagger}{\bar d}_{p;a}\right]
+V\cdot\frac{3\Delta^2}{2G_c}\ . \ \ 
\eeq


Next, let us derive and solve the gap equation for CFL phase with $F_3=F_8=0$ to obtain $\Delta$. 
The vacuum state is written as $\ket{\Phi}$ which is a vacuum with respect to 
the quasi-particle operators $d_{p;a}$ and ${\bar d}_{p;a}$: 
\beq\label{22}
d_{p;a}\ket{\Phi}={\bar d}_{p;a}\ket{\Phi}=0\ .
\eeq
Thus, the thermodynamic potential $\Phi_0$ can be expressed as 
\beq\label{23}
\Phi_0&=&\frac{1}{V}\cdot \bra{\Phi}H_{\rm eff}\ket{\Phi}\nonumber\\
&=&\frac{1}{2V}\sum_{p\ (\epsilon_p>\mu)}\left[9{\bar \epsilon}_p-\sqrt{{\bar \epsilon}_p^2+4\Delta^2}-8\sqrt{{\bar \epsilon}_p^2+\Delta^2}\right]
\nonumber\\
& &
+\frac{1}{2V}\sum_{p\ (\epsilon_p<\mu)}\left[9{\bar \epsilon}_p-\sqrt{{\bar \epsilon}_p^2+4\Delta^2}-8\sqrt{{\bar \epsilon}_p^2+\Delta^2}\right]
+\frac{3\Delta^2}{2G_c}
\ .\quad
\eeq

The gap equation, $\partial\Phi_0/\partial \Delta=0$, can be expressed as 
\beq\label{24}
\Delta\left(\frac{3}{G_c}-\int^{\Lambda}\frac{d^3{\mib p}}{(2\pi)^3}
\left(\frac{4}{\sqrt{{\bar \epsilon}_p^2+4\Delta^2}}+\frac{8}{\sqrt{{\bar \epsilon}_p^2+\Delta^2}}\right)\right)=0\ ,
\eeq
where $\Lambda$ represents a three-momentum cutoff. 
Here, the helicity $\eta=\pm 1$ is considered which gives a factor 2. 
Namely, the above gap equation with $\Delta \neq 0$ is written as
\beq\label{25}
1=\frac{2G_c}{3\pi^2}\int_0^{\Lambda}dp\ p^2\left(\frac{1}{\sqrt{(p-\mu)^2+4\Delta^2}}+\frac{2}{\sqrt{(p-\mu)^2+\Delta^2}}\right)\ , 
\eeq
where $p=|{\mib p}|$. 
Solving (\ref{25}) with respect to $\Delta$ and substituting its solution into (\ref{23}), 
the thermodynamic potential (\ref{23}) is obtained:
\beq\label{26}
\Phi_0=\frac{1}{2\pi^2}\int_0^{\Lambda}dp\ p^2\left(
9(p-\mu)-\sqrt{(p-\mu)^2+4\Delta^2}-8\sqrt{(p-\mu)^2+\Delta^2}\right)+\frac{3\Delta^2}{2G_c}\ . 
\nonumber\\
& &
\eeq
The right-hand sides of Eqs.(\ref{25}) and (\ref{26}) can be analytically expressed by using the following formulae:
\beq\label{27}
\int_0^{\Lambda}dp\frac{p^2}{\sqrt{(p-\mu)^2+c\Delta^2}}
&=&
\left[\frac{p+3\mu}{2}\sqrt{p^2-2\mu p+\mu^2+c\Delta^2}\right]_0^{\Lambda}
\nonumber\\
& &\!\!\!
+\left[\left(\!\mu^2\!-\!\frac{c\Delta^2}{2}\!\!\right)\ln\!\left|2p-2\mu+2\sqrt{p^2-2\mu p+\mu^2+c\Delta^2}\right|\right]_0^{\Lambda}\!\! , 
\nonumber\\
\int_0^{\Lambda}dp\ p^2\sqrt{(p-\mu)^2+c\Delta^2}
&=&\left[\frac{3p+5\mu}{12}(p^2-2\mu p+\mu^2+c\Delta^2)^{3/2}\right]_0^{\Lambda}\nonumber\\
&+&\left[\frac{4\mu^2-c\Delta^2}{8}\left((p-\mu)\sqrt{p^2-2\mu p+\mu^2+c\Delta^2}\right.\right.\nonumber\\
&+&
c\Delta^2\ln\left|2p-2\mu+2\sqrt{p^2-2\mu p+\mu^2+c\Delta^2}\right|\biggl)\Biggl]_0^{\Lambda}\! .\ \ 
\eeq

Of course, if $\Delta=0$, the thermodynamic potential is given by 
\beq\label{28}
\Phi_0(\Delta=0)=-\frac{3\mu^4}{8\pi^2}\ . 
\eeq

\subsection{Spin polarized phase without color superconducting gap $\Delta$}

In this subsection, we derive the thermodynamic potential with $\Delta=0$. 
In this case, it is only necessary to diagonalize the Hamiltonian matrix (\ref{6}), namely,
\beq\label{29}
\kappa
&=&\left(
\begin{array}{cccc}
p & 0 & 0 & 0 \\
0 & p & 0 & 0 \\
0 & 0 & -p & 0 \\
0 & 0 & 0 & -p \\
\end{array}
\right)+
\frac{{\cal F}_\tau}{p}
\left(
\begin{array}{cccc}
0  & \sqrt{p_1^2+p_2^2} & -p_3  & 0 \\
\sqrt{p_1^2+p_2^2} & 0 & 0 & p_3 \\
-p_3 & 0 & 0 & \sqrt{p_1^2+p_2^2} \\
0 & p_3 & \sqrt{p_1^2+p_2^2} & 0 
\end{array}
\right)
\nonumber\\
&=&\left(
\begin{array}{cccc}
q & e_{\tau} & -g_{\tau} & 0 \\
e_{\tau} & q & 0 & g_{\tau} \\
-g_{\tau} & 0 & -q & e_{\tau} \\
0 & g_{\tau} & e_{\tau} & -q \\
\end{array}
\right)\ ,
\eeq
where $q=p$, $e_{\tau}={\cal F_\tau}\sqrt{p_1^2+p_2^2}/p$ and $g_{\tau}={\cal F}_{\tau}p_3/p$. 
The eigenvalues of $\kappa$ are easily obtained as 
\beq\label{30}
\pm\epsilon_{{\mib p}\tau}^{(\eta)}=\pm\sqrt{g_{\tau}^2+(e_{\tau}+\eta q)^2}=\pm\sqrt{p_3^2+\left({\cal F}_{\tau}+\eta\sqrt{p_1^2+p_2^2}\right)^2}\ , 
\eeq
where ${\cal F}_{\tau}$ are defined in (\ref{5}) or ({\ref{2}}). 
Since the Hamiltonian is diagonalized, the thermodynamic potential with $\Delta=0$, which is written as $\Phi_F$, can be easily obtained as 
\beq\label{31}
& &\Phi_F=3\frac{1}{V}\sum_{{\mib p}}\sum_{\eta=\pm}\sum_{\tau=u,d,s}(\epsilon_{{\mib p}\tau}^{(\eta)}-\mu)\theta(\mu-\epsilon_{{\mib p}\tau}^{(\eta)})
+\frac{1}{2G}(F_3^2+F_8^2)\ ,
\eeq
where the factor 3 represents the degree of freedom of color.
Here, the sum with respect to the momentum should be replaced to the integration of momentum:
$(1/V)\cdot\sum_{\mib p}\rightarrow \int d^3{\mib p}$. 
Hereafter, we assume $|{\cal F}_{\tau}|<\mu$ because we are interested in the phase transition from CFL phase to SP phase. 
In the case $0\leq {\cal F}_{\tau}<\mu$, the ranges of integration are obtained as
\beq\label{32}
& &{\rm for}\ \ \eta=-1, \nonumber\\
& &
\ \ 
0\leq p_{\perp}=\sqrt{p_1^2+p_2^2}\leq {\cal F}_{\tau}+\mu\ , \ \ 
-\sqrt{\mu^2-({\cal F}_{\tau}-p_{\perp})^2}\leq p_3 \leq \sqrt{\mu^2-({\cal F}_{\tau}-p_{\perp})^2}\ , \quad\nonumber\\
& &{\rm for}\ \ \eta=1, \nonumber\\
& &
\ \  
0\leq p_{\perp}=\sqrt{p_1^2+p_2^2}\leq \mu-{\cal F}_{\tau}\ , \ \ 
-\sqrt{\mu^2-({\cal F}_{\tau}+p_{\perp})^2}\leq p_3 \leq \sqrt{\mu^2-({\cal F}_{\tau}+p_{\perp})^2}\ . \nonumber\\
& & 
\eeq 
In the case $-\mu<{\cal F}_{\tau}\leq 0$, the ranges of integration are obtained as
\beq\label{33}
& &{\rm for}\ \ \eta=-1, \nonumber\\
& &
\ \ 
0\leq p_{\perp}=\sqrt{p_1^2+p_2^2}\leq \mu-|{\cal F}_{\tau}|\ , \ \ 
-\sqrt{\mu^2-(|{\cal F}_{\tau}|+p_{\perp})^2}\leq p_3 \leq \sqrt{\mu^2-(|{\cal F}_{\tau}|+p_{\perp})^2}\ , \quad\nonumber\\
& &{\rm for}\ \ \eta=1, \nonumber\\
& &
\ \ 
0\leq p_{\perp}=\sqrt{p_1^2+p_2^2}\leq |{\cal F}_{\tau}|+\mu\ , \ \ 
-\sqrt{\mu^2-(|{\cal F}_{\tau}|-p_{\perp})^2}\leq p_3 \leq \sqrt{\mu^2-(|{\cal F}_{\tau}|-p_{\perp})^2}\ . \!\!\!\!
\nonumber\\
& &
\eeq
Regardless of the sign of ${\cal F}_{\tau}$, positive or negative, 
${\cal F}_{\tau}$ is regarded as the absolute value of ${\cal F}_{\tau}$ since we consider both $\eta=1$ and $\eta=-1$. 
Thus, from (\ref{31}) and (\ref{32}) or (\ref{33}), 
the thermodynamic potential can be expressed  
as is similar to Eq.(29) in Ref. \cite{oursPTEP}.
\beq\label{34}
\Phi_F&=&
\frac{1}{2G}(F_3^2+F_8^2)\nonumber\\
& &+\frac{3}{2\pi^2}\sum_{\tau}
\Biggl[
\int_0^{\mu-|{\cal F}_{\tau}|}dp_{\perp}\ p_{\perp}\int_{0}^{\sqrt{\mu^2-(|{\cal F}_{\tau}|+p_{\perp})^2}}
dp_3\left(\sqrt{p_3^2+(|{\cal F}_{\tau}|+p_{\perp})^2}-\mu\right)
\nonumber\\
& &\qquad\qquad
+\int_0^{\mu+|{\cal F}_{\tau}|}dp_{\perp}\ p_{\perp}\int_{0}^{\sqrt{\mu^2-(|{\cal F}_{\tau}|-p_{\perp})^2}}
dp_3\left(\sqrt{p_3^2+(|{\cal F}_{\tau}|-p_{\perp})^2}-\mu\right)\Biggl]
\nonumber\\
&=&\frac{1}{2G}(F_3^2+F_8^2)\nonumber\\
& &-\frac{1}{2\pi^2}\sum_{\tau}
\biggl[
\frac{\sqrt{\mu^2-{\cal F}_{\tau}^2}}{4}(3{\cal F}_{\tau}^2\mu+2\mu^3)+{\cal F}_{\tau}\mu^3\arctan\left(\frac{{\cal F}_{\tau}}{\sqrt{\mu^2-{\cal F}_{\tau}^2}}\right)
\nonumber\\
& &\qquad\qquad\qquad\qquad\qquad\qquad\qquad\qquad
-\frac{{\cal F}_{\tau}^4}{8}\ln\left(\frac{(\mu+\sqrt{\mu^2-{\cal F}_{\tau}^2})^2}{{\cal F}_{\tau}^2}\right)\biggl]\ , 
\eeq 
where ${\cal F}_u=F_3+F_8/\sqrt{3},\ {\cal F}_d=-F_3+F_8/\sqrt{3}$ and ${\cal F}_s=-2F_8/\sqrt{3}$. 
The simultaneous gap equations with respect to $F_3$ and $F_8$ are obtained through $\partial \Phi_F/\partial F_3=0$ and 
$\partial \Phi_F/\partial F_8=0$ as 
\beq\label{35}
\frac{\partial \Phi_F}{\partial F_k}&=&\frac{F_k}{G}
-\frac{1}{2\pi^2}\sum_{\tau=u,d,s}\Biggl(
2{\cal F}_{\tau}\mu\sqrt{\mu^2-{\cal F}_{\tau}^2}+\mu^3\arctan\left(\frac{{\cal F}_{\tau}}{\sqrt{\mu^2-{\cal F}_{\tau}^2}}\right)
\nonumber\\
& &\qquad\qquad\qquad\qquad
-\frac{{\cal F}_{\tau}^3}{2}
\ln\left(\frac{(\mu+\sqrt{\mu^2-{\cal F}_{\tau}^2})^2}{{\cal F}_{\tau}^2}\right)\Biggl)\cdot
\frac{\partial {\cal F}_{\tau}}{\partial F_k}\nonumber\\
&=&0\ ,
\eeq
where $k=3$ or $8$ and $\partial {\cal F}_u/\partial F_3=1,\ \partial {\cal F}_d/\partial F_3=-1,\ \partial {\cal F}_s/\partial F_3=0,\ 
\partial {\cal F}_u/\partial F_8=1/\sqrt{3},\ \partial {\cal F}_d/\partial F_8=1/\sqrt{3}$ and $\partial {\cal F}_s/\partial F_8=-2/\sqrt{3}$.  
Inserting the solutions of above simultaneous gap equations into the thermodynamic potential $\Phi_F$, 
we can estimate $\Phi_F$, which should be compared with $\Phi_0$ in (\ref{26}). 
By comparing (\ref{34}) with (\ref{26}), the realized phase is determined.

\setcounter{equation}{0}

\section{Numerical results}

Let us calculate the thermodynamic potential $\Phi_0$ with $F_k=0$ in (\ref{23}) and $\Phi_F$ with $\Delta=0$ in 
(\ref{31}) or (\ref{34}) numerically. 
If $\Phi_F<\Phi_0$, the SP phase is realized. 
However, in rather smaller $\mu$, the CFL phase should be realized. 
The numerical results are summarized in Table {\ref{table1}}. 
%
\begin{table}[b]
\caption{Numerical results
}
\label{table1}
\centering
\begin{tabular}{c|cc|ccc}
\hline
$\mu/{\rm GeV}$ & $\Delta/{\rm GeV}$ & $\Phi_0/{\rm GeV}^4$ & $F_3/{\rm GeV}$ & 
$F_8/{\rm GeV}$ & $\Phi_F/{\rm GeV}^{4}$  \\ 
\hline
0.40 & 0.042640 & $-0.0020336$ & 0 & 0 & $-0.0019454$  
\\
0.41 & 0.045135 & $-0.0022509$ & 0.048509 & 0.028007 & $-0.0021481$ 
\\
0.42 & 0.047509 & $-0.0024845$ & 0.098544 & 0.056895 & $-0.0023753$ 
\\
0.43 & 0.049748 & $-0.0027352$ & 0.136983 & 0.079087 & $-0.0026327$  
\\
0.44 & 0.051841 & $-0.0030034$ & 0.170908 & 0.098674 & $-0.0029245$  
\\
0.45 & 0.053774 & $-0.0032897$ & 0.202345 & 0.116824 & $-0.0032542$ 
\\
0.4558 & 0.054817 & $-0.0034643$ & 0.219828 & 0.126918 & $-0.0034643$ 
\\
0.46 & 0.055536 & $-0.0035947$ & 0.232234 & 0.134080 & $-0.0036256$  
\\
0.47 & 0.057116 & $-0.0039189$ & 0.261119 & 0.150757 & $-0.0040423$  
\\
0.48 & 0.058502 & $-0.0042627$ & 0.289365 & 0.167065 & $-0.0045084$ 
\\
0.49 & 0.059681 & $-0.0046267$ & 0.317255 & 0.183167 & $-0.0050277$ 
\\
0.50 & 0.060640 & $-0.0050113$ & 0.345041 & 0.199210 & $-0.0056046$  
\\
\hline
\end{tabular}
\end{table}
%
We adopt the parameters as 
$\Lambda=0.631$ GeV, $G_c=6.6$ GeV${}^{-2}$ and $G=20$ GeV${}^{-2}$.
As is seen in Table \ref{table1}, in the region of $\mu \leq 0.45$ GeV, 
$\Phi_0 < \Phi_F$ is satisfied and the CFL phase is realized. 
At $\mu=0.4558$ GeV $(=\mu_c)$, $\Phi_0\approx \Phi_F$ is satisfied. 
Thus, the phase transition may occur. 
In the region of $\mu \geq 0.46$ GeV, $\Phi_0>\Phi_F$ is satisfied and the realized phase is 
the SP phase.

As for the solutions of the simultaneous gap equations in (\ref{35}), 
it seems that a relation $F_3 \approx \sqrt{3}F_8$ may be satisfied. 
If $F_3=\sqrt{3}F_8$, the relation ${\cal F}_d={\cal F}_s(=-2F_3/3)$ is derived. 
In this case, ${\cal F}_u=4F_3/3(=-2{\cal F}_d)$. 
As for another case, for example, $F_3=0$ and $F_8\neq 0$, 
then ${\cal F}_u={\cal F}_d$ and ${\cal F}_s=-2{\cal F}_u$ are satisfied. 
So, another local minimum may be obtained at $(F_3=0,\ F_8)$. 

\begin{figure}[t]
\begin{center}
\includegraphics[height=5.5cm]{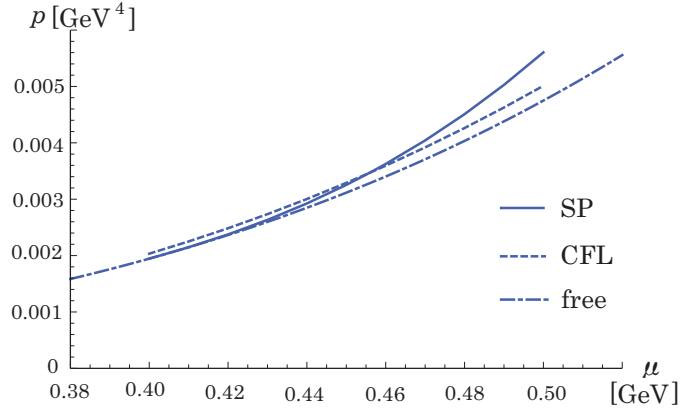}
\caption{The pressures for $(\Delta\neq 0, F_3=F_8=0)$ (CFL), $(\Delta=0, F_3\neq 0, F_8\neq 0)$ (SP) and free quark gas (free) are 
depicted as functions of the quark chemical potential $\mu$.
}
\label{fig:fig1}
\end{center}
\end{figure}

Of course, the pressure $p_A$ can be expressed by the thermodynamic potential $\Phi_A$ through the thermodynamical relation: 
\beq\label{37}
p_A=-\Phi_A\ .
\eeq
In Fig.\ref{fig:fig1}, the pressures of the CFL phase and the SP phase are depicted as a function 
of the chemical potential $\mu$ with a comparison of that of the free quark matter.
In $\mu < \mu_c$ ($\mu>\mu_c$), the pressure of CFL phase is 
larger (smaller) than that of SP phase. 
Thus, the realized phase is CFL (SP) phase.

Also, the quark number density $\rho_q$ can be derived from the thermodynamic potential by thermodynamical relation:
\beq\label{36}
\rho_q=-\frac{\partial \Phi}{\partial \mu}\ . 
\eeq
The baryon number density $\rho_B$ can be expressed as $\rho_B=\rho_q/3$. 
In Table \ref{table2}, we summarize the baryon density and baryon density divided by the normal nuclear density $\rho_0=0.17$ fm${}^{-3}$.

%

%
\begin{table}[b]
\caption{Numerical results
}
\label{table2}
\centering
\begin{tabular}{c|ccc|ccc}
\hline
{\scriptsize $\mu/{\rm GeV}$} & {\scriptsize $\rho_B(F=0)/{\rm fm}{}^{-3}$} & \!\!\!\!{\scriptsize $\rho_B(F=0)/\rho_0$}\!\!\!\! & {\scriptsize $p_0/{\rm GeV}{}^4$} & 
{\scriptsize $\rho_B(\Delta=0)/{\rm fm}{}^{-3}$} &\!\!\!\! {\scriptsize $\rho_B(\Delta=0)/\rho_0$} \!\!\!\!& {\scriptsize $p_F/{\rm GeV}{}^4$} \\ 
\hline
0.40 & 2.73681 & 5.36630 & 0.0020336 &  &
\\
0.41 & 2.94752 & 5.77946 & 0.0022509 & 2.79121 & 5.47297 & 0.0021481 
\\
0.42 & 3.16606 & 6.20796 & 0.0024845 & 3.16045 & 6.19697 & 0.0023753 
\\
0.43 & 3.39217 & 6.65131 & 0.0027352 & 3.58384 & 7.02714 & 0.0016327 
\\
0.44 & 3.62561 & 7.10904 & 0.0030034 & 4.05631 & 7.95354 & 0.0029245 
\\
0.45 & 3.86614 & 7.58067 & 0.0032897 & 4.57688 & 8.97428 & 0.0032542 
\\
0.4558 & 4.00881 & 7.8604 & 0.0034643 & 4.90099 & 9.60978 & 0.0034643 
\\
0.46 & 4.11353 & 8.06574 & 0.0035947 & 5.14599 & 10.0902 & 0.0036256
\\
0.47 & 4.36753 & 8.56379 & 0.0039189 & 5.76482 & 11.3036 & 0.0040423
\\
0.48 & 4.62793 & 9.07438 & 0.0042627 & 6.43516 & 12.618 & 0.0045084
\\
0.49 & 4.8945 & 9.59706 & 0.0046267 & 7.1594 & 14.038 & 0.0050277
\\
0.50 & 5.16702 & 10.1314 & 0.0050113 & 7.94077 & 15.5701 & 0.0056046 
\\
\hline
\end{tabular}
\end{table}
%

Finally, let us estimate the spin polarization. 
Here, we consider the helicity instead of spin. 
The quark number density for the flavor $\tau$ and helicity $\eta$ can be derived as 
\beq\label{38}
n_{\tau}^{(\eta)}=3\int\frac{d^3{\mib p}}{(2\pi)^3}\theta(\mu-\epsilon_{{\mib p}\tau}^{(\eta)})\ , 
\eeq
where the factor 3 means color degree of freedom.
First, in the case of ${\cal F}_{\tau}$ being positive and $0<{\cal F}_{\tau}<\mu$, 
the particle number density with helicity $\pm$, $n_{\tau >}^{(\pm)}$, can be calculated as 
\beq\label{39}
n_{\tau>}^{(+)}&=&\frac{3}{2\pi^2}\int_0^{\mu-{\cal F}_{\tau}}dp_{\perp}p_{\perp}\int_0^{\sqrt{\mu^2-({\cal F}_{\tau}+p_{\perp})^2}}dp_3\nonumber\\
&=&\frac{1}{4\pi^2}\left[
\sqrt{\mu^2-{\cal F}_{\tau}^2}({\cal F}_{\tau}^2+2\mu^2)+3{\cal F}_{\tau}\mu^2\arctan \frac{{\cal F}_{\tau}}{\sqrt{\mu^2-{\cal F}_{\tau}^2}}
-\frac{3{\cal F}_{\tau}\mu^2\pi}{2}\right]\ ,\nonumber\\
n_{\tau>}^{(-)}&=&\frac{3}{2\pi^2}\int_0^{\mu+{\cal F}_{\tau}}dp_{\perp}p_{\perp}\int_0^{\sqrt{\mu^2-({\cal F}_{\tau}-p_{\perp})^2}}dp_3\nonumber\\
&=&\frac{1}{4\pi^2}\left[
\sqrt{\mu^2-{\cal F}_{\tau}^2}({\cal F}_{\tau}^2+2\mu^2)+3{\cal F}_{\tau}\mu^2\arctan \frac{{\cal F}_{\tau}}{\sqrt{\mu^2-{\cal F}_{\tau}^2}}
+\frac{3{\cal F}_{\tau}\mu^2\pi}{2}\right]\ . \ \ 
\eeq
On the other hand, in the case $-\mu < {\cal F}_{\tau}<0$, the particle number density $n_{\tau <}^{(\eta)}$ is calculated as  
\beq\label{40}
n_{\tau<}^{(+)}&=&\frac{3}{2\pi^2}\int_0^{\mu+|{\cal F}_{\tau}|}dp_{\perp}p_{\perp}\int_0^{\sqrt{\mu^2-(|{\cal F}_{\tau}|-p_{\perp})^2}}dp_3\nonumber\\
&=&\frac{1}{4\pi^2}\left[
\sqrt{\mu^2-|{\cal F}_{\tau}|^2}(|{\cal F}_{\tau}|^2+2\mu^2)+3|{\cal F}_{\tau}|\mu^2\arctan \frac{|{\cal F}_{\tau}|}{\sqrt{\mu^2-|{\cal F}_{\tau}|^2}}
+\frac{3|{\cal F}_{\tau}|\mu^2\pi}{2}\right]\ , 
\nonumber\\
n_{\tau<}^{(-)}&=&\frac{3}{2\pi^2}\int_0^{\mu-|{\cal F}_{\tau}|}dp_{\perp}p_{\perp}\int_0^{\sqrt{\mu^2-(|{\cal F}_{\tau}|+p_{\perp})^2}}dp_3\nonumber\\
&=&\frac{1}{4\pi^2}\left[
\sqrt{\mu^2-|{\cal F}_{\tau}|^2}(|{\cal F}_{\tau}|^2+2\mu^2)+3|{\cal F}_{\tau}|\mu^2\arctan \frac{|{\cal F}_{\tau}|}{\sqrt{\mu^2-|{\cal F}_{\tau}|^2}}
-\frac{3|{\cal F}_{\tau}|\mu^2\pi}{2}\right]\ . \nonumber\\
& & 
\eeq
For the case ${\cal F}_\tau>\mu$ , there is no contribution to the spin polarization from $\epsilon_{{\mib p}\tau}^{(+)}$ due to integration range. 
Thus, we obtain only $n_{\tau >}^{(-)}$:
\beq\label{41}
n_{\tau>}^{(-)}&=&3\int\frac{d^3{\mib p}}{(2\pi)^3}\theta(\mu-\epsilon_{{\mib p}\tau}^{(-)})
=\frac{3\mu^2}{4\pi}{\cal F}_{\tau}\ . 
\eeq
It should be here noted that the fermi surface has a form of torus in the case ${\cal F}_{\tau} > \mu$, 
namely, $(\sqrt{p_1^2+p_2^2}-{\cal F}_{\tau})^2+p_3^2=\mu^2$,
whose volume is obtained as $2\pi {\cal F}_{\tau}\times \pi \mu^2$ where 
${\cal F}_{\tau}$ and $\mu$ correspond to the major and minor radius of torus. 
For the case ${\cal F}_{\tau}<-\mu$, as is similar to Eq.(\ref{41}), we obtain the following:
\beq\label{42}
n_{\tau<}^{(+)}=\frac{3\mu^2}{4\pi}|{\cal F}_{\tau}|\ . 
\eeq

\begin{figure}[b]
\begin{center}
\includegraphics[height=3cm]{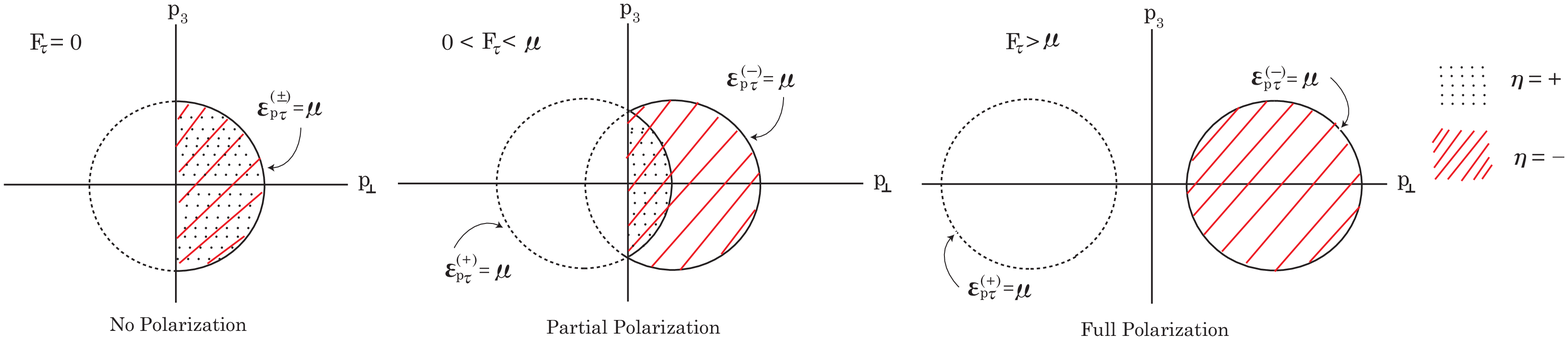}
\caption{The fermi surface is depicted for ${\cal F}_\tau>0$. 
In this case, the minus helicity is dominated. 
}
\label{fig:fig3}
\end{center}
\end{figure}

Next, let us consider the case $|{\cal F}_{\tau}|<\mu$. 
The spin polarization per unit volume is obtained by the difference between the quark number with the plus helicity and that with minus helicity, 
as is shown in 
Fig.\ref{fig:fig3}.  
Further, we assume that ${\cal F}_u>0$, ${\cal F}_d<0$ and ${\cal F}_s<0$. 
The spin polarization of each flavor, ${\cal S}_{\tau}$, can be expressed as 
\beq\label{43}
{\cal S}_u/(\hbar/2)=n_{u>}^{(+)}-n_{u>}^{(-)}=-\frac{3}{4\pi}{\cal F}_u\mu^2\ .
\eeq
Similarly, 
\beq\label{44}
{\cal S}_d/(\hbar/2)=n_{d<}^{(+)}-n_{d<}^{(-)}=\frac{3}{4\pi}|{\cal F}_d|\mu^2\ , \quad
{\cal S}_s/(\hbar/2)=n_{s<}^{(+)}-n_{s<}^{(-)}=\frac{3}{4\pi}|{\cal F}_s|\mu^2\ . \ \ \ \ 
\eeq
Of course, the total spin polarization ${\cal S}$ is written as 
\beq\label{45}
{\cal S}={\cal S}_u+{\cal S}_d+{\cal S}_s\ . 
\eeq
%
\begin{figure}[t]
\begin{center}
\includegraphics[height=5.5cm]{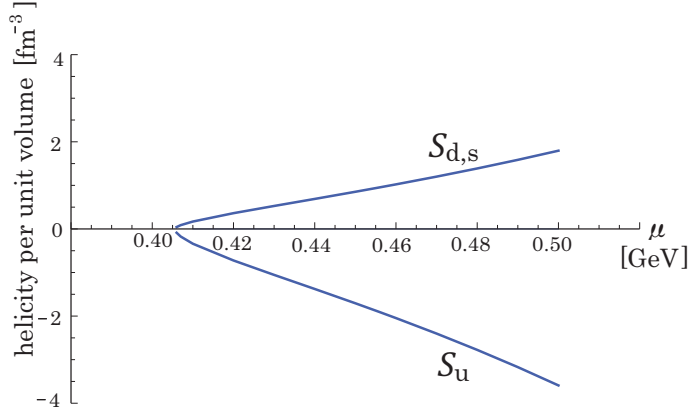}
\caption{The spin polarization of each flavor is depicted as a function of the quark chemical potential $\mu$. 
}
\label{fig:fig4}
\end{center}
\end{figure}
%
Figure \ref{fig:fig4} shows the numerical results. 
For $d$- and $s$-quarks, the spin polarization has almost the same magnitude. 
On the other hand, for $u$-quark, the spin polarization is opposite to $d$- and $s$-quarks.
The total spin polarization is nearly equal to be zero. 
The reason is as follows: 
From the numerical calculation, the condensate $F_3$ and $F_8$ satisfy the relation $F_3\approx \sqrt{3}F_8$ which makes 
the thermodynamic potential be minimum. 
Under this relation, ${\cal F}_u=-2{\cal F}_d$ and ${\cal F}_d={\cal F}_s$ are satisfied. 
Thus, from (\ref{43}) $\sim$ (\ref{45}), ${\cal S}=0$ is derived.

\setcounter{equation}{0}

\section{The order of phase transition and the second order perturbation with respect to $V_{\rm SP}$}

In order to investigate the order of phase transition from CFL phase to SP phase, 
we treat the interaction term $V_{\rm SP}$ in the perturbation theory on the vacuum of the CFL phase. 
We neglect the contribution of negative-energy particle represented by $({\tilde c}_{p\tau\alpha}, {\tilde c}_{p\tau\alpha}^{\dagger})$. 
Thus, in $V_{\rm SP}$ in (\ref{7}), it is enough to pick up only the following term: 
\beq\label{46}
& &H_1=\sum_{{\mib p}\eta\tau\alpha}{\cal F}_{\tau}\frac{\sqrt{p_1^2+p_2^2}}{|{\mib p}|}c_{{\mib p}\eta\tau\alpha}^{\dagger}c_{{\mib p}-\eta\tau\alpha}
=\sum_{{\mib p}\eta\tau\alpha}e_{\tau}c_{{\mib p}\eta\tau\alpha}^{\dagger}c_{{\mib p}-\eta\tau\alpha}\ , 
\nonumber\\
& &e_{\tau}\equiv
{\cal F}_{\tau}\frac{\sqrt{p_1^2+p_2^2}}{|{\mib p}|}\ . 
\eeq
The correction of the first-order perturbation with respect to $H_1$, 
namely $\bra{\Phi}H_1\ket{\Phi}$, vanishes because 
the helicity $\eta$ is different from each other like  
$\bra{\Phi}d_{{\mib p}\eta;a}d_{{\mib p}-\eta;a}^{\dagger}\ket{\Phi}=0$, 
which is led from 
$\bra{\Phi}c_{{\mib p}\eta\tau\alpha}^{\dagger}c_{{\mib p}-\eta\tau\alpha}\ket{\Phi}$. 
Thus, we need to calculate the second-order perturbation with respect to $H_1$. 
The correction of energy can be expressed as  
\beq\label{47}
E_{\rm corr}=\sum_{i}\frac{\bra{\Phi}H_1\ket{i}\bra{i}H_1\ket{\Phi}}{E_{0}-E_i}\ , 
\eeq
where $\ket{i}$, $E_0$ and $E_i$ represent the excited state, the vacuum energy and 
the excited energies, respectively. 
The term $H_1$ includes $c_{{\mib p}\eta\alpha\tau}^{\dagger} c_{{\mib p}-\eta\alpha\tau}$. 
For example, in the case $\epsilon_p>\mu$ with $\tau=u$, 
the necessary term can be expressed in terms of the quasiparticle operators $(d_{{\mib p}\eta;i}, d_{{\mib p}\eta;i}^{\dagger})$ such as 
\beq\label{48}
\sum_{\alpha}c_{{\mib p}\eta\alpha u}^{\dagger} c_{{\mib p}-\eta\alpha u}
&=&c_{{\mib p}\eta 1u}^{\dagger} c_{{\mib p}-\eta 1u}
+c_{{\mib p}\eta 2u}^{\dagger} c_{{\mib p}-\eta 2u}
+c_{{\mib p}\eta 3u}^{\dagger} c_{{\mib p}-\eta 3u}
\nonumber\\
&=&
[x_p^{(1)}d_{{\mib p}\eta;1}^{\dagger}+x_p^{(2)}d_{{\mib p}\eta;2}^{\dagger}
+x_p^{(3)}d_{{\mib p}\eta;3}^{\dagger}
\nonumber\\
& &
-(y_p^{(1)}d_{-{\mib p}\eta;1}+y_p^{(2)}d_{-{\mib p}\eta;2}
+y_p^{(3)}d_{-{\mib p}\eta;3})]\nonumber\\
& &\times [x_p^{(1)}d_{{\mib p}-\eta;1}+x_p^{(2)}d_{{\mib p}-\eta;2}
+x_p^{(3)}d_{{\mib p}-\eta;3}
\nonumber\\
& &\qquad
-(y_p^{(1)}d_{-{\mib p}-\eta;1}^{\dagger}+y_p^{(2)}d_{-{\mib p}-\eta;2}^{\dagger}
+y_p^{(3)}d_{-{\mib p}-\eta;3}^{\dagger})]
\nonumber\\
& &+(X_pd_{{\mib p}\eta 2u}^{\dagger}-Y_pd_{-{\mib p}\eta 1d})
(X_pd_{{\mib p}-\eta 2u}-Y_pd_{-{\mib p}-\eta 1d})
\nonumber\\
& &+(X_pd_{{\mib p}\eta 3u}^{\dagger}-Y_pd_{-{\mib p}\eta 1s})
(X_pd_{{\mib p}-\eta 3u}-Y_pd_{-{\mib p}-\eta 1s})\ .
\eeq
Therefore, the intermediate states $\ket{i}$ are needed: 
\beq\label{49}
\ket{i}=\ket{{\mib p}\eta ab}\equiv d_{{\mib p}\eta;a}^{\dagger}d_{-{\mib p}-\eta;b}^{\dagger}
\ket{\Phi}\ . \qquad
(\bra{i}=\bra{\Phi}d_{-{\mib p}-\eta;b}d_{{\mib p}\eta;a}
)
\eeq 
Here, we remember $d_{{\mib p}\eta;4}=d_{{\mib p}\eta 2u}$ and so on.
We summarize the necessary pieces to calculate the second order perturbative correction of energy for $\epsilon_p >\mu$ in Appendix B.
Thus, for $\epsilon_p>\mu$, the second order perturbative correction, 
$E_{\rm corr}^{>}$, for the 
energy can be expressed as 
\beq
E_{\rm corr}^{>}&=&
\sum_{{\mib p}\eta ab}\frac{
\bra{\Phi}H_1\ket{{\mib p}\eta ab}\bra{{\mib p}\eta ab}H_1\ket{\Phi}}
{E_0-E_{{\mib p}\eta ab}}\nonumber\\
&=&
-\sum_{{\mib p}\eta(\epsilon>\mu)}
\biggl\{
\frac{(e_u+e_d+e_s)^2(x_p^{(1)}y_p^{(1)})^2}{2\sqrt{{\bar \epsilon}^2+4\Delta^2}}
\nonumber\\
& &\qquad\qquad\ \ 
+\frac{1}{\sqrt{{\bar \epsilon}^2+4\Delta^2}+\sqrt{{\bar \epsilon}^2+\Delta^2}}
\biggl[
(e_u-e_d)^2(x_p^{(2)}y_p^{(1)}+x_p^{(1)}y_p^{(2)})^2
\nonumber\\
& &\qquad\qquad\qquad\qquad\qquad\qquad\qquad\qquad
+(e_u+e_d-2e_s)^2(x_p^{(1)}y_p^{(3)}+x_p^{(3)}y_p^{(1)})^2\biggl]
\nonumber\\
& &\qquad\qquad
+\frac{1}{2\sqrt{{\bar \epsilon}^2+\Delta^2}}
\biggl[
(e_u+e_d)^2(x_p^{(2)}y_p^{(2)})^2+(e_u-e_d)^2(x_p^{(2)}y_p^{(3)}+x_p^{(3)}y_p^{(2)})^2
\nonumber\\
& &\qquad\qquad\qquad\qquad\qquad
+(e_u+e_d+4e_s)^2(x_p^{(3)}y_p^{(3)})^2\biggl]
\nonumber\\
& &\qquad\qquad
+\frac{1}{2\sqrt{{\bar \epsilon}^2+\Delta^2}}
\left[
(e_u+e_d)^2+(e_u+e_s)^2+(e_d+e_s)^2\right](X_pY_p)^2\biggl\}
\ .
\eeq

As for the case $\epsilon_p < \mu$, the same results are derived. 
The intermediate states are adopted as
\beq\label{52}
\ket{{\mib p}\eta \ovl{ab}}={\bar d}_{-{\mib p}\eta;a}^{\dagger}
{\bar d}_{{\mib p}-\eta;b}^{\dagger}\ket{\Phi}\ .
\eeq
Then, the structure of operators in $H_1$ for $\epsilon_p <\mu$ 
is the same as that of the case for $\epsilon_p > \mu$. 
Thus, the correction energy, $E_{\rm corr}^{<}$, for $\epsilon_p < \mu$ has the same form:  
\beq\label{53}
E_{\rm cror}^{<}&=&
\sum_{{\mib p}\eta ab}\frac{
\bra{\Phi}H_1\ket{{\mib p}\eta \ovl{ab}}\bra{{\mib p}\eta \ovl{ab}}H_1\ket{\Phi}}
{E_0-E_{{\mib p}\eta \ovl{ab}}}\nonumber\\
&=&
-\sum_{{\mib p}\eta(\epsilon<\mu)}
\biggl\{
\frac{(e_u+e_d+e_s)^2({\bar x}_p^{(1)}{\bar y}_p^{(1)})^2}{2\sqrt{{\bar \epsilon}^2+4\Delta^2}}
\nonumber\\
& &\qquad\qquad
+\frac{1}{\sqrt{{\bar \epsilon}^2+4\Delta^2}+\sqrt{{\bar \epsilon}^2+\Delta^2}}
\biggl[
(e_u-e_d)^2({\bar x}_p^{(2)}{\bar y}_p^{(1)}+{\bar x}_p^{(1)}{\bar y}_p^{(2)})^2
\nonumber\\
& &\qquad\qquad\qquad\qquad\qquad\qquad\qquad\qquad
+(e_u+e_d-2e_s)^2({\bar x}_p^{(1)}{\bar y}_p^{(3)}+{\bar x}_p^{(3)}{\bar y}_p^{(1)})^2\biggl]
\nonumber\\
& &\qquad\qquad
+\frac{1}{2\sqrt{{\bar \epsilon}^2+\Delta^2}}
\biggl[
(e_u+e_d)^2({\bar x}_p^{(2)}{\bar y}_p^{(2)})^2+(e_u-e_d)^2({\bar x}_p^{(2)}{\bar y}_p^{(3)}+{\bar x}_p^{(3)}{\bar y}_p^{(2)})^2
\nonumber\\
& &\qquad\qquad\qquad\qquad\qquad
+(e_u+e_d+4e_s)^2({\bar x}_p^{(3)}{\bar y}_p^{(3)})^2\biggl]
\nonumber\\
& &\qquad\qquad
+\frac{1}{2\sqrt{{\bar \epsilon}^2+\Delta^2}}
\left[
(e_u+e_d)^2+(e_u+e_s)^2+(e_d+e_s)^2\right]({\bar X}_p{\bar Y}_p)^2\biggl\}
\ .
\eeq
Finally, we obtain the correction energy by the second-order perturbation as 
\beq\label{54}
E_{\rm corr}=E_{\rm corr}^{>}+E_{\rm corr}^{<}\ . 
\eeq 
Here, by using $x_p^{(a)}$, $y_p^{(a)}$, $X_p$, $Y_p$ in (\ref{15}),  ${\bar x}_p^{(a)}$, ${\bar y}_p^{(a)}$, ${\bar X}_p$, ${\bar Y}_p$ in (\ref{16}), 
$e_{\tau}$ in (\ref{46}) and ${\cal F}_{\tau}$ in (\ref{2}), and by substituting the above quantities  into $E_{\rm corr}$, we obtain 
\beq\label{55}
E_{\rm corr}
&=&
-2\left(\sum_{{\mib p}\ (\epsilon>\mu)}+\sum_{{\mib p}\ (\epsilon<\mu)}\right)
\frac{p_1^2+p_2^2}{|{\mib p}|^2}
\biggl\{
\frac{1}{\sqrt{{\bar \epsilon}^2+4\Delta^2}+\sqrt{{\bar \epsilon}^2+\Delta^2}}
\nonumber\\
& &\qquad\qquad\qquad\qquad\qquad\qquad
\times \frac{1}{3}
\left(1-\frac{{\bar \epsilon}^2+2\Delta^2}{\sqrt{{\bar \epsilon}^2+\Delta^2}\sqrt{{\bar \epsilon}^2+
4\Delta^2}}\right)\cdot\left(F_3^2+F_8^2\right) 
\nonumber\\
& &\qquad\qquad\quad\qquad\qquad\quad\qquad
+\frac{1}{2\sqrt{{\bar \epsilon}^2+\Delta^2}}\cdot\frac{\Delta^2}{{\bar \epsilon}^2+\Delta^2}
\left(\frac{5}{6}F_3^2+\frac{19}{12}F_8^2\right)\biggl\}\ . \qquad \ 
\eeq
Then, the thermodynamic potential $\Phi$ can be obtained up to the second order of $V_{\rm SP}$, namely, 
up to the second order of $F_k$ as 
\beq\label{56}
\Phi=\Phi_0+\frac{1}{V}\cdot E_{\rm corr}+\frac{1}{2G}(F_3^2+F_8^2)\ . 
\eeq

%
\begin{table}[t]
\caption{Numerical results
}
\label{table3}
\centering
\begin{tabular}{c|cc}
\hline
$\mu/{\rm GeV}$ &  $c_3+1/(2G)$ & $c_8+1/(2G)$ \\ 
\hline
0.40  & 0.015734 & 0.0076297 
\\
0.41 & 0.015304 & 0.0068182 
\\
0.42 & 0.014873 & 0.0060047 
\\
0.43 & 0.014427 & 0.0051920 
\\
0.44 & 0.014015 & 0.0043828 
\\
0.45 & 0.013592 & 0.0035803 
\\
0.4558 & 0.0133487 & 0.0031934 
\\
0.46 & 0.013174 & 0.0027882
\\
0.47 & 0.012765 & 0.0020104
\\
0.48 & 0.012367 & 0.0012519
\\
0.49 & 0.011982 & 0.00051816
\\
0.50 & 0.116142 & $-0.00018406$ 
\\
\hline
\end{tabular}
\end{table}
%

The order of the phase transition form CFL to SP phases is determined through (\ref{55}). 
Namely, from (\ref{55}) and (\ref{56}), for CFL phase, but with small $F_3$ and/or $F_8$, 
the thermodynamic potential $\Phi$ is obtained: 
\beq\label{57}
\Phi
&=&\Phi_0+\left(c_3+\frac{1}{2G}\right)F_3^2+\left(c_8+\frac{1}{2G}\right)F_8^2
\eeq
up to the second order of $F_k$. 
Here, $c_3$ and $c_8$ are expressed as 
\beq\label{58}
c_3&=&
-\frac{1}{\pi^2}\int_0^{\Lambda} dp_3\int_0^{\sqrt{\Lambda^2-p_3^2}}dp_{\perp}\ p_{\perp}
\nonumber\\
& &\qquad
\times
\biggl[\frac{p_{\perp}^2}{|{\mib p}|^2}\biggl\{\frac{1}{\sqrt{|{\mib p}|^2-2\mu|{\mib p}|+\mu^2+4\Delta^2}+\sqrt{|{\mib p}|^2-2\mu|{\mib p}|+\mu^2+\Delta^2}}
\nonumber\\
& &\qquad\qquad\qquad
\times \frac{1}{3}\left(1-\frac{|{\mib p}|^2-2\mu|{\mib p}|+\mu^2+2\Delta^2}{
\sqrt{|{\mib p}|^2-2\mu|{\mib p}|+\mu^2+4\Delta^2}\sqrt{|{\mib p}|^2-2\mu|{\mib p}|+\mu^2+\Delta^2}}\right)
\nonumber\\
& &\qquad\qquad\qquad\qquad\qquad\qquad\qquad\qquad
+\frac{\Delta^2}{2(|{\mib p}|^2-2\mu|{\mib p}|+\mu^2+\Delta^2)^{3/2}}\cdot \frac{5}{6}\biggl\}\biggl]\ , \nonumber\\
c_8&=&
-\frac{1}{\pi^2}\int_0^{\Lambda} dp_3\int_0^{\sqrt{\Lambda^2-p_3^2}}dp_{\perp}\ p_{\perp}
\nonumber\\
& &\qquad
\times
\biggl[\frac{p_{\perp}^2}{|{\mib p}|^2}\biggl\{\frac{1}{\sqrt{|{\mib p}|^2-2\mu|{\mib p}|+\mu^2+4\Delta^2}+\sqrt{|{\mib p}|^2-2\mu|{\mib p}|+\mu^2+\Delta^2}}
\nonumber\\
& &\qquad\qquad\qquad
\times \frac{1}{3}\left(1-\frac{|{\mib p}|^2-2\mu|{\mib p}|+\mu^2+2\Delta^2}{
\sqrt{|{\mib p}|^2-2\mu|{\mib p}|+\mu^2+4\Delta^2}\sqrt{|{\mib p}|^2-2\mu|{\mib p}|+\mu^2+\Delta^2}}\right)
\nonumber\\
& &\qquad\qquad\qquad\qquad\qquad\qquad\qquad\qquad
+\frac{\Delta^2}{2(|{\mib p}|^2-2\mu|{\mib p}|+\mu^2+\Delta^2)^{3/2}}\cdot \frac{19}{12}\biggl\}\biggl]\  .
\nonumber\\
& &  
\eeq
If $c_3+1/(2G)>0$ and $c_8+1/(2G)>0$, the phase transition from CFL to SP phases is of the first order 
because $F_3=F_8=0$ always gives a local minimum of the thermodynamic potential $\Phi_0$. 
On the other hand, if $c_3+1/(2G)>0$ and $c_8+1/(2G)<0$ and vice versa, or $c_3+1/(2G)<0$ and $c_8+1/(2G)<0$, the phase transition from CFL to SP 
phases is maybe of the second order. 
As is seen in Table \ref{table3}, in the region of $\mu \leq 0.49$ GeV, especially, 
at $\mu=\mu_c\ (=0.4558$ GeV), the coefficients $c_3+1/(2G)$ and $c_8+1/(2G)$ are positive. 
Thus, the phase transition may be of the first order.

\begin{figure}[t]
\begin{center}
\includegraphics[height=5.5cm]{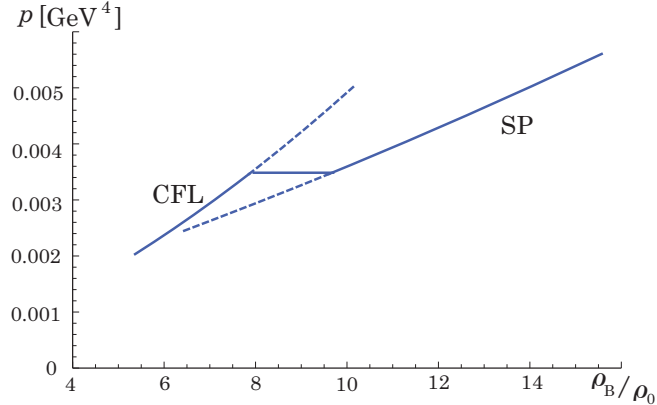}
\caption{The pressure is depicted as a function of the baryon number density divided by the normal nuclear density. 
The solid curve represents the realized phase.
}
\label{fig:fig2}
\end{center}
\end{figure}

From the above consideration, in Fig.\ref{fig:fig2}, the pressure is depicted as a function of the baryon number density divided by normal 
nuclear density, which has already given in Table \ref{table2}. 
The realized phase is represented by solid curve.


\section{Summary and concluding remarks}

In this paper, it has been shown that the quark spin polarization for each flavor may occur in three-flavor case at high baryon density, 
which leads to the quark spin polarized phase, against the color-flavor locked phase due to 
the four-point tensor-type interaction between quarks in the Nambu-Jona-Lasinio model. 
In a certain region of quark chemical potential, the CFL phase is favorable energetically. 
However, as the quark chemical potential increases, the phase transition from CFL phase to 
spin polarized phase occurs. 
In our theoretical model, the phase transition occurs around the quark chemical potential being around 0.45 GeV under a certain parameter set.
Based on the CFL phase, we have treated the tensor-type interaction term between quarks as a perturbation one. 
As a results, it has been shown that the phase transition may be of the first order up to the second-order perturbation.

In this paper, it has been shown that, owing to the four-point tensor-type interaction, the spin polarization 
may occur in the NJL model. 
The tensor-type interaction may come from the two-gluon exchange term between quarks in QCD. 
It is interesting to clarify the origin of the tensor-type interaction. 
In this paper, the total spin polarization is not realized, while the spin polarization with 
respect to each flavor actually occurs. 
It may be important to introduce the quark-mass splitting, namely, the strange 
quark mass should be taken into account, while we have ignored it in this paper. 
Further, if the chiral symmetry is explicitly broken, namely, the quark masses are not zero even in the chiral symmetric phase, 
the spin polarization originated from the pseudovector-type four-point interaction between quarks may exist. 
One of next interesting problems may be to investigate the interplay between the spin polarization from tensor-type interaction and 
that from the pseudovector-type interaction. 
As for a realistic calculation of the inner core of neutron stars or quark stars, 
the charge neutrality condition is important \cite{AR} and then, 
the chemical potential of each quark flavor should be introduced. 
This is one of future problems we consider, while 
we expect that the spin polarized phase may appear by the effects developed in 
this paper. 
Further, it is interesting to investigate the origin of strong magnetic field \cite{IF}, for example, in the core of neutron stars. 
It is suggested that, in general, the spin polarization leads to the ferromagnetization.  
In the core of neutron stars, it is expected that there exists the strong magnetic field. 
Then, if the high density quark matter is realized in the inner core of neutron stars or quark stars, 
the spin polarization may occur as is shown in this paper. 
Thus, one direction of investigations in high density quark matter is to understand whether the quark ferromagnetization is 
realized or not in the quark spin polarized phase. 
This will be one thing to solve.

\section*{Acknowledgment}

One of the authors (Y.T.) would like to express his sincere thanks to 
Professor\break
J. da Provid\^encia and Professor C. Provid\^encia, two of co-authors of this paper, 
for their warm hospitality during his visit to Coimbra in spring of 2014. 
One of the authors (Y.T.) 
is partially supported by the Grants-in-Aid of the Scientific Research 
(No.23540311, No.26400277) from the Ministry of Education, Culture, Sports, Science and 
Technology in Japan.

\vspace{-0cm}

\appendix

\section{Mean field Hamiltonian without spin polarization in terms of the quasiparticle operators}

The mean field Hamiltonian, $H_{\rm eff}$, in (\ref{9}) can be expressed in terms of the quasiparticle operators 
$d_{p:i},\ d_{p;i}^{\dagger},\ {\bar d}_{{\bar p};i}$ and ${\bar d}_{{\bar p};i}^{\dagger}$. 
The result is as follows:
\beq\label{a1}
H_{\rm eff}&=&H_>+H_< +V\cdot\frac{3\Delta^2}{2G_c}\ ,\nonumber\\
H_>&=&
\frac{1}{2}\sum_{p\ (\epsilon_p>\mu)}(4{\bar \epsilon}_p^2Y_p^2+4\Delta X_pY_p\phi_p)\times 3\nonumber\\
& &+\frac{1}{2}\sum_{p\ (\epsilon_p>\mu)}\left[
{\bar \epsilon}_p\cdot 6(y_p^{(1)2}+y_p^{(2)2}+y_p^{(3)2})+2\Delta (-6x_p^{(1)}y_p^{(1)}+2x_p^{(2)}y_p^{(2)}+6x_p^{(3)}y_p^{(3)})\phi_p\right]
\nonumber\\
& &+\sum_{p\ (\epsilon_p>\mu)}\biggl\{
\left[{\bar \epsilon}_p\cdot 3(x_p^{(1)2}-y_p^{(1)2})+12\Delta x_p^{(1)}y_p^{(1)}\phi_p\right]d_{p;1}^{\dagger}d_{p;1}\nonumber\\
& &\qquad\qquad
+\left[{\bar \epsilon}_p\cdot 2(x_p^{(2)2}-y_p^{(2)2})-4\Delta x_p^{(2)}y_p^{(2)}\phi_p\right]d_{p;2}^{\dagger}d_{p;2}\nonumber\\
& &\qquad\qquad
+\left[{\bar \epsilon}_p\cdot 6(x_p^{(3)2}-y_p^{(3)2})-12\Delta x_p^{(3)}y_p^{(3)}\phi_p\right]d_{p;3}^{\dagger}d_{p;3}\nonumber\\
& &\qquad\qquad\quad
+\sum_{a=4}^{9}\left[{\bar \epsilon}_p\cdot (X_p^{2}-Y_p^{2})-2\Delta X_p Y_p\phi_p\right]d_{p;a}^{\dagger}d_{p;a}\biggl\}\nonumber\\
& &+\sum_{p\ (\epsilon_p>\mu)}\biggl\{
\left[-6{\bar \epsilon}_p x_p^{(1)}y_p^{(1)}+6\Delta (x_p^{(1)2}-y_p^{(1)2})\phi_p\right](d_{p;1}^{\dagger}d_{{\bar p};1}^{\dagger}+d_{{\bar p};1}d_{p1})\nonumber\\
& &\qquad\qquad
+\left[-4{\bar \epsilon}_p x_p^{(2)}y_p^{(2)}-2\Delta (x_p^{(2)2}-y_p^{(2)2})\phi_p\right](d_{p;2}^{\dagger}d_{{\bar p};2}^{\dagger}+d_{{\bar p};2}d_{p2})\nonumber\\
& &\qquad\qquad
+\left[-12{\bar \epsilon}_p x_p^{(3)}y_p^{(3)}-6\Delta (x_p^{(3)2}-y_p^{(3)2})\phi_p\right](d_{p;3}^{\dagger}d_{{\bar p};3}^{\dagger}+d_{{\bar p};3}d_{p3})\nonumber\\
& &\qquad\qquad\quad
+\left[-2{\bar \epsilon}_p X_p Y_p-\Delta (X_p^2-Y_p^2)\phi_p\right]
\nonumber\\
& &\qquad\qquad\qquad
\times
(d_{p2u}^{\dagger}d_{{\bar p}1d}^{\dagger}+d_{p1d}^{\dagger}d_{{\bar p}2u}^{\dagger}+
d_{{\bar p}1d}d_{p2u}+d_{{\bar p}2u}d_{p1d}
+d_{p3u}^{\dagger}d_{{\bar p}1s}^{\dagger}+d_{p1s}^{\dagger}d_{{\bar p}3u}^{\dagger}\nonumber\\
& &\qquad\qquad\qquad\ \ 
+d_{{\bar p}1s}d_{p3u}+d_{{\bar p}3u}d_{p1s}
+d_{p3d}^{\dagger}d_{{\bar p}2s}^{\dagger}+d_{p2s}^{\dagger}d_{{\bar p}3d}^{\dagger}+d_{{\bar p}2s}d_{p3d}+d_{{\bar p}3d}d_{p2s})
\biggl\}, \nonumber\\
H_<&=&
\frac{1}{2}\sum_{p\ (\epsilon_p<\mu)}(4{\bar \epsilon}_p^2{\bar X}_p^2-4\Delta {\bar X}_p {\bar Y}_p\phi_p)\times 3\nonumber\\
& &+\frac{1}{2}\sum_{p\ (\epsilon_p<\mu)}\left[
{\bar \epsilon}_p\cdot 2(3{\bar x}_p^{(1)2}+2{\bar x}_p^{(2)2}+6{\bar x}_p^{(3)2})+2\Delta (6{\bar x}_p^{(1)}y_p^{(1)}
-2{\bar x}_p^{(2)}{\bar y}_p^{(2)}-6{\bar x}_p^{(3)}{\bar y}_p^{(3)})\phi_p\right]
\nonumber\\
& &+\sum_{p\ (\epsilon_p<\mu)}\biggl\{
\left[{\bar \epsilon}_p\cdot 3({\bar y}_p^{(1)2}-{\bar x}_p^{(1)2})-12\Delta {\bar x}_p^{(1)}{\bar y}_p^{(1)}\phi_p\right]{\bar d}_{p;1}^{\dagger}{\bar d}_{p;1}\nonumber\\
& &\qquad\qquad
+\left[{\bar \epsilon}_p\cdot 2({\bar y}_p^{(2)2}-{\bar x}_p^{(2)2})+4\Delta {\bar x}_p^{(2)}{\bar y}_p^{(2)}\phi_p\right]{\bar d}_{p;2}^{\dagger}{\bar d}_{p;2}\nonumber\\
& &\qquad\qquad
+\left[{\bar \epsilon}_p\cdot 6({\bar y}_p^{(3)2}-{\bar x}_p^{(3)2})+12\Delta {\bar x}_p^{(3)}{\bar y}_p^{(3)}\phi_p\right]{\bar d}_{p;3}^{\dagger}{\bar d}_{p;3}\nonumber
\\
& &\qquad\qquad\quad
+\sum_{a=4}^{9}\left[{\bar \epsilon}_p\cdot ({\bar Y}_p^{2}-{\bar X}_p^{2})+2\Delta {\bar X}_p {\bar Y}_p\phi_p\right]{\bar d}_{p;a}^{\dagger}{\bar d}_{p;a}\biggl\}\nonumber\\
& &+\sum_{p\ (\epsilon_p<\mu)}\biggl\{
\left[-6{\bar \epsilon}_p {\bar x_p}^{(1)}{\bar y}_p^{(1)}+6\Delta ({\bar x}_p^{(1)2}-{\bar y}_p^{(1)2})\phi_p\right]
({\bar d}_{{\bar p};1}{\bar d}_{p;1}+{\bar d}_{p;1}^{\dagger}{\bar d}_{{\bar p};1}^{\dagger})\nonumber\\
& &\qquad\qquad
+\left[-4{\bar \epsilon}_p {\bar x}_p^{(2)}{\bar y}_p^{(2)}-2\Delta ({\bar x}_p^{(2)2}-{\bar y}_p^{(2)2})\phi_p\right]
({\bar d}_{{\bar p};2}{\bar d}_{p;2}+{\bar d}_{p;2}^{\dagger}{\bar d}_{{\bar p};2}^{\dagger})\nonumber\\
& &\qquad\qquad
+\left[-12{\bar \epsilon}_p {\bar x}_p^{(3)}{\bar y}_p^{(3)}-6\Delta ({\bar x}_p^{(3)2}-{\bar y}_p^{(3)2})\phi_p\right]
({\bar d}_{{\bar p};3}{\bar d}_{p;3}+{\bar d}_{p;3}^{\dagger}{\bar d}_{{\bar p};3}^{\dagger})\nonumber\\
& &\qquad\qquad\quad
+\left[-2{\bar \epsilon}_p {\bar X}_p {\bar Y}_p-\Delta ({\bar X}_p^2-{\bar Y}_p^2)\phi_p\right]\nonumber\\
& &\qquad\qquad\qquad
\times
({\bar d}_{{\bar p}2u}{\bar d}_{p1d}+{\bar d}_{{\bar p}1d}{\bar d}_{p2u}+{\bar d}_{p1d}^{\dagger}{\bar d}_{{\bar p}2u}^{\dagger}+{\bar d}_{p2u}^{\dagger}{\bar d}_{{\bar p}1d}^{\dagger}
+{\bar d}_{{\bar p}3u}{\bar d}_{p1s}+{\bar d}_{{\bar p}1s}{\bar d}_{p3u}
\nonumber\\
& &\qquad\qquad\qquad\ \ 
+{\bar d}_{p1s}^{\dagger}{\bar d}_{{\bar p}3u}^{\dagger}+{\bar d}_{p3u}^{\dagger}{\bar d}_{{\bar p}1s}^{\dagger}
+{\bar d}_{{\bar p}3d}{\bar d}_{p2s}+{\bar d}_{{\bar p}2s}{\bar d}_{p3d}+{\bar d}_{p2s}^{\dagger}{\bar d}_{{\bar p}3d}^{\dagger}+{\bar d}_{p3d}^{\dagger}{\bar d}_{{\bar p}2s}^{\dagger})
\biggl\}. \qquad\ \ 
\eeq
Substituting $x_p^{(a)}$ and so on, we obtain the simple form in (\ref{20}).

\section{Necessary matrix elements in the second order perturbation theory}

We collect the necessary pieces to calculate the second-order correction of energy, (\ref{47}). 
\beq\label{b1}
\bra{\Phi}H_1\ket{{\mib p}\eta 11}\bra{{\mib p}\eta 11}H_1\ket{\Phi}
& &\nonumber\\
=
\sum_{{\mib p}'\eta'}\bra{\Phi}(-x_{p'}^{(1)}y_{p'}^{(1)})
(e_u+e_d& &\!\!\!\!\!\!
+e_s)
d_{-{\mib p}'\eta';1}d_{{\mib p}'-\eta';1}\cdot
d_{{\mib p}\eta;1}^{\dagger}d_{-{\mib p}-\eta;1}^{\dagger}\ket{\Phi}\nonumber\\
\times  
\sum_{{\mib p}'\eta'}\bra{\Phi}d_{-{\mib p}-\eta;1}d_{{\mib p}\eta;1}
(-& &\!\!\!\!\!\!x_{p'}^{(1)}y_{p'}^{(1)})(e_u+e_d+e_s)
d_{{\mib p}'\eta';1}^{\dagger}d_{-{\mib p}'-\eta';1}^{\dagger}\ket{\Phi}
\nonumber\\
&=&(x_p^{(1)}y_p^{(1)})^2(e_u+e_d+e_s)^2\ , \nonumber\\
E_0-E_{11}&=&-2\sqrt{{\bar \epsilon}^2+4\Delta^2}\ , \nonumber\\
\bra{\Phi}H_1\ket{{\mib p}\eta 12}\bra{{\mib p}\eta 12}H_1\ket{\Phi}
&=&(x_p^{(2)}y_p^{(1)}+x_p^{(1)}y_p^{(2)})^2(-e_u+e_d)^2\ , \nonumber\\
E_0-E_{12}&=&-\sqrt{{\bar \epsilon}^2+4\Delta^2}-\sqrt{{\bar \epsilon}^2+\Delta^2}\ , \nonumber\\
\bra{\Phi}H_1\ket{{\mib p}\eta 13}\bra{{\mib p}\eta 13}H_1\ket{\Phi}
&=&(x_p^{(1)}y_p^{(3)}+x_p^{(3)}y_p^{(1)})^2(-e_u-e_d+2e_s)^2\ , \nonumber\\
E_0-E_{13}&=&-\sqrt{{\bar \epsilon}^2+4\Delta^2}-\sqrt{{\bar \epsilon}^2+\Delta^2}\ , \nonumber\\
\bra{\Phi}H_1\ket{{\mib p}\eta 22}\bra{{\mib p}\eta 22}H_1\ket{\Phi}
&=&(x_p^{(2)}y_p^{(2)})^2(e_u+e_d)^2\ , \nonumber\\
E_0-E_{22}&=&-2\sqrt{{\bar \epsilon}^2+\Delta^2}\ , \nonumber\\
\bra{\Phi}H_1\ket{{\mib p}\eta 23}\bra{{\mib p}\eta 23}H_1\ket{\Phi}
&=&(x_p^{(2)}y_p^{(3)}+x_p^{(3)}y_p^{(2)})^2(-e_u+e_d)^2\ , \nonumber\\
E_0-E_{23}&=&-2\sqrt{{\bar \epsilon}^2+4\Delta^2}\ , \nonumber\\
\bra{\Phi}H_1\ket{{\mib p}\eta 33}\bra{{\mib p}\eta 33}H_1\ket{\Phi}
&=&(x_p^{(3)}y_p^{(3)})^2(e_u+e_d+4e_s)^2\ , \nonumber\\
E_0-E_{33}&=&-2\sqrt{{\bar \epsilon}^2+4\Delta^2}\ , \nonumber
\\
\bra{\Phi}H_1\ket{{\mib p}\eta 45}\bra{{\mib p}\eta 45}H_1\ket{\Phi}
&=&(X_p Y_p)^2(e_u+e_d)^2\ , \nonumber\\
E_0-E_{45}&=&-2\sqrt{{\bar \epsilon}^2+4\Delta^2}\ , \nonumber\\
\bra{\Phi}H_1\ket{{\mib p}\eta 67}\bra{{\mib p}\eta 67}H_1\ket{\Phi}
&=&(X_p Y_p)^2(e_u+e_s)^2\ , \nonumber
\\
E_0-E_{67}&=&-2\sqrt{{\bar \epsilon}^2+4\Delta^2}\ , \nonumber
\\
\bra{\Phi}H_1\ket{{\mib p}\eta 89}\bra{{\mib p}\eta 89}H_1\ket{\Phi}
&=&(X_p Y_p)^2(e_d+e_s)^2\ , \nonumber\\
E_0-E_{89}&=&-2\sqrt{{\bar \epsilon}^2+4\Delta^2}\ .
\eeq


%


\end{document}